\documentclass[a4paper,11pt]{article}
\pdfoutput=1 
\usepackage{jinstpub} 

\usepackage{subcaption}
\usepackage{xcolor}
\usepackage{epstopdf}
\graphicspath{
    {.}
}

\DeclareMathOperator\sign{sign}
\DeclareMathOperator\cov{cov}
\newcommand{\ee}{\mathrm{e}}
\newcommand{\GeV}{\textrm{GeV}}

\title{Gaussian Process-based calculation of look-elsewhere trials factor}

\author{V. Ananiev,}
\author{A. L. Read}
\affiliation{Department of Physics, University of Oslo, Boks 1072 Blindern, Oslo, NO-0316, Norway}

\emailAdd{a.l.read@fys.uio.no}

\abstract{%
In high-energy physics it is a recurring challenge to efficiently and precisely (enough) calculate
the global significance of, e.g., a potential new resonance.
We propose a new method that models the significance in the search region as a Gaussian Process.
The kernel of the Gaussian Process is approximated with a covariance matrix and is calculated
with a carefully designed set of background-only data sets,
comparable in number to the random background-only data sets used in a typical analysis that relies on the average upcrossings of the significance. The trials factor for both low and moderate significances
can subsequently be calculated to the desired precision with a computationally inexpensive random sampling of the Gaussian Process.
In addition, once the covariance of the Gaussian Process is determined, the average number of upcrossings can be computed analytically.
In our work we give some highlights of the analytic calculation
and also discuss some peculiarities of the trials factor estimation on a finite grid.
We illustrate the method with studies of three complementary statistical models.}

\arxivnumber{2206.12328}

\begin{document}

\maketitle
\flushbottom

\section{Introduction}
In a typical search for a new particle or resonance in a high-energy physics experiment, the statistical model consists of a relatively narrow signal peak on top of a broad background in the invariant mass spectrum. In general, the location of the signal is not known a priori so the likelihood is maximized for signal masses and amplitudes in the ``search region'' of mass. In some models the width of the signal is not uniquely specified, leading to a 2-dimensional search region. The background model has typically one or more nuisance parameters that describe the form and normalization of the background distribution. The signal parameters such as width and mass, however, are not defined under the background hypothesis. In this case, the significance at the location of maximum likelihood of the signal plus background hypothesis (local significance) should be reduced to account for random fluctuations of the background that can occur anywhere in the search region. The corrected significance is called ``global significance'', and the ratio between the corresponding global and local p-values or tail probabilities, which is a function of the observed local significance, is called the ``trials factor'' (TF). The whole situation is often referred to in high energy physics as the ``look-elsewhere effect''.

The most straightforward way to estimate the trials factor is to analyse a large ensemble of Monte Carlo data sets (toy data sets or ``toys'') generated from the background model and to count how many times the significance exceeds a local significance level. However, Monte Carlo toys are usually expensive to analyse in terms of computational resources.

\begin{figure}[ht]
    \centering
    \includegraphics[width=0.48\linewidth]{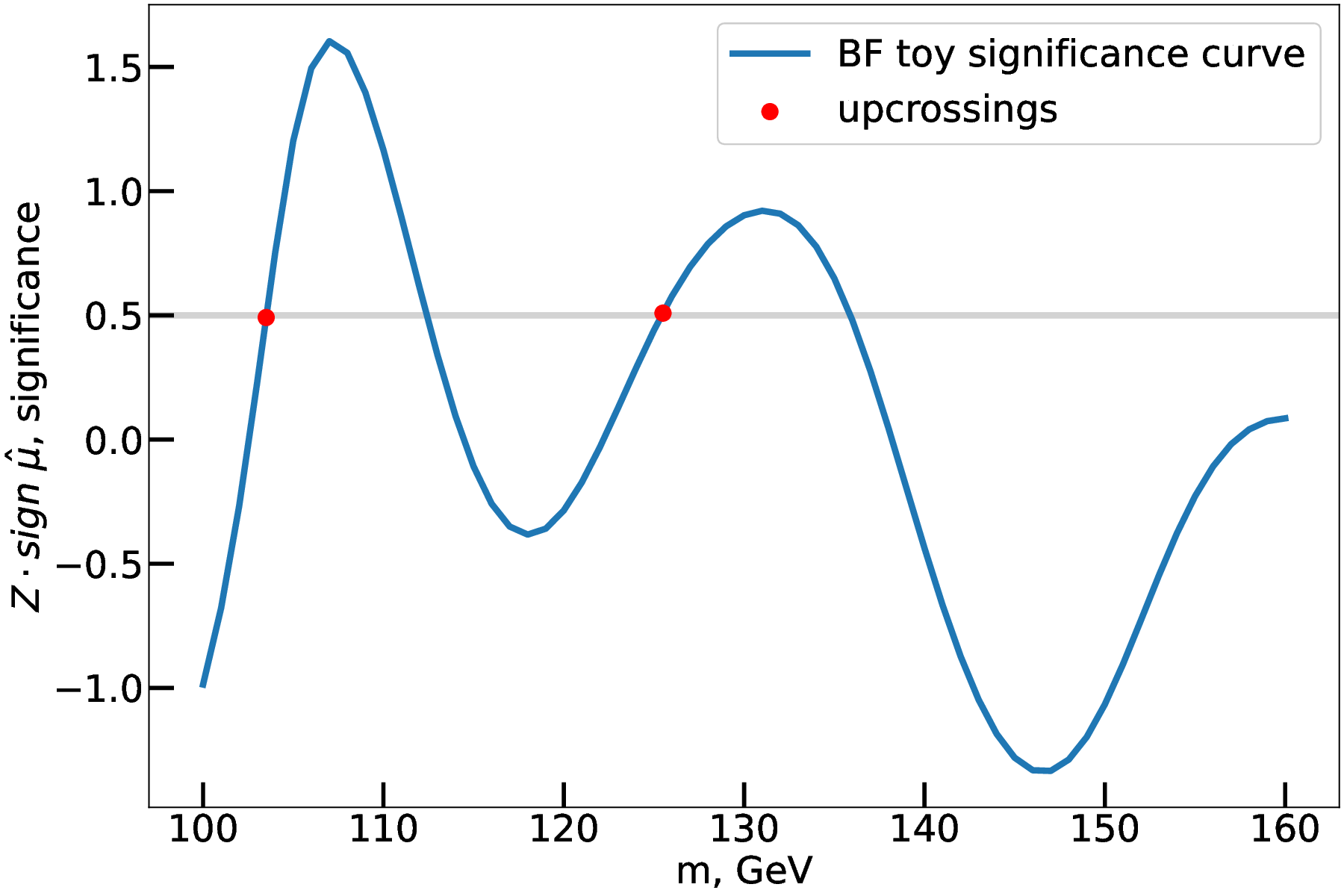}
    \caption{%
An example of a significance field (blue curve) for a likelihood ratio scan of the invariant mass
of the Higgs boson in a study of brute force (BF) toys.
Red dots denote locations of the upcrossings of threshold $0.5$ by this significance curve.}\label{fig:hyy-upcrossings}
\end{figure}

Gross and Vitells, in their studies of the look-elsewhere effect~\cite{gross2010,vitells2011},
suggested a way to estimate the upper bound on the trials factor,
which for large local significances $\gtrsim 3\sigma$ approaches the true value.
Their method sets the upper bound to the trials factor based on the average Euler characteristic of the $\chi^2$ field,
a field that emerges from the signal nuisance parameter scan.
In the one-dimensional case the average Euler characteristic can be estimated from
the average number of upcrossings (figure~\ref{fig:hyy-upcrossings})
counted at some reference level of local significance and then propagated to higher levels.
This method, however, can be very conservative when applied to lower significances.
In this work, we argue that a relatively small number of carefully designed background toys can be used efficiently to provide more precise and
less conservative estimates for the trials factor,
as well as more precise estimates for low-significance upcrossings.

We focus on models with a test statistic that follows a $\chi^2$ distribution with 1 degree of freedom.
This allows us to turn the $\chi^2$ field into a significance field (figure~\ref{fig:hyy-upcrossings})
that we then propose in section~\ref{sec:method} to approximate with a Gaussian Process (GP) whose kernel,
approximated with a covariance matrix calculated with the carefully designed toys.
The resulting GP allows us to efficiently sample approximate Monte Carlo toys and,
therefore, to quickly evaluate the trials factor via an approximate brute force method.

It is worth mentioning that the idea of using the GP in relation to hypothesis testing when a nuisance parameter is present only under the alternative is not new
and was studied by Davies~\cite{davies77}.
We, however, build on top of this idea with a suggestion of a practical way to estimate the GP kernel.

As described in section~\ref{subsec:upcross_adhoc}, once the covariance of the GP is determined, the average number of upcrossings can, in fact, be computed analytically. We also discuss some peculiarities of the trials factor estimation on a finite grid in section~\ref{subsec:sneaks}.

We demonstrate our method applied to the background template model used by Gross and Vitells (GV) in their study (section~\ref{subsec:example_gv}) as well as exponential background models with 1-dimensional (section~\ref{subsec:example_hyy}) and 2-dimensional (section~\ref{subsec:example_hyy2d}) search regions inspired by searches for a resonance due to $H\rightarrow\gamma\gamma$ decays in proton-proton collisions at the Large Hadron Collider~\cite{PhysRevLett.108.111803,CMSHyy}.

The Jupyter notebook and the data that were used to produce figures for this paper are publicly available at Zenodo~\cite{zenodo-data}.
The Python package that we developed and used to produce and fit the data sets, which is also extensively used in the notebook for visualizations,
is also publicly available~\cite{gitlab-sigcorr} and includes the documentation~\cite{gitlab-sigcorr-docs} with recipes of how to reproduce the results we present in this work.

\pagebreak
\section{Method}\label{sec:method}
In resonance searches the signal model has a nuisance parameter $M$ that is not present under the background-only hypothesis and that denotes the set of points that constitute the search region. The search region itself may be one-dimensional, for a case when the mass of a hypothetical particle is unknown, or many dimensional, for example 2-dimensional, when the decay width is also not specified uniquely. We elaborate on the 2-D search region in section~\ref{subsec:example_hyy2d}.

We focus on the case where the signal model has 1 free parameter
(e.g.\@ signal amplitude or strength $\mu$, that we assume is fitted unconstrained\footnotemark[1])
for each point in $M$, therefore the profile likelihood test statistic
\footnotetext[1]{We need both positive and negative values of the best-fit signal strength $\hat{\mu}$
to be able to reconstruct the sign of the significance and, therefore, both sides of its distribution.
This is a crucial step required to treat the significance as a GP,
which is at the core of the method suggested in this paper.
We, however, argue that unconstrained fits generally are not an issue for analyses of large samples,
while analyses of small samples can benefit from brute force calculations of the TF
with no significant penalty in performance.}
\begin{align}
t(M) = -2\ln \lambda(M),\ \lambda(M)= \frac{\mathcal{L}(0,\hat{\hat{\theta}})}{\mathcal{L}(\hat\mu,M,\hat\theta)}
\end{align}
follows a $\chi^2$ distribution with 1 degree of freedom in the large sample limit~\cite{wilks1938}. Here $\theta$ is one or more additional nuisance parameters that maximize the likelihood under the background ($\hat{\hat{\theta}}$) or signal plus background ($\hat{\theta}$) hypotheses.

We want to construct a standard normal random variable from $t(M)$. For large samples:
\begin{align}
    &\sqrt{t(M)} \sim \sqrt{\chi^2_1} = \left| N(0, 1) \right|.
\end{align}
To restore the sign of the standard normal distribution we use the sign of the signal strength $\hat{\mu}$, which is known to be normally distributed around $0$ in the absence of signal, in the large sample limit as a maximum likelihood estimator.

We then construct the normally distributed significance $Z(M)$ as follows:
\begin{align}
    Z(M) &= \sqrt{t(M)}\sign{\hat{\mu}}. \label{eq:signed_lrt}
\end{align}

To compute the trials factor we need to determine, under the background hypothesis $H_0: \mu = 0$, the probability that $Z(M)$ exceeds
the local significance observed in the data for any $M$ in the search region.

We propose to approximate the ensemble of curves $Z(M)$ with a family of samples from some Gaussian Process. Such an approximation increases the performance of the sampling without introducing significant bias. Properties of $Z$-curves (significance curves) are reflected in the GP mean, that should be set to $0$, and the kernel $K$ with $K(M_i, M_i) = 1$, i.e., the kernel must have 1 on the diagonal.

The crucial step of our proposal is to estimate $\hat{\Sigma}$, which is the covariance matrix, an approximation of the GP kernel evaluated on the grid specific to the problem.
Properties of the kernel $K$, and subsequently of the covariance matrix $\hat{\Sigma}$, can be formally expressed as follows:
\begin{align}\label{eq:sigcor_definition}
    \hat{\Sigma}_{ij} &= \cov\left[Z(M_i), Z(M_j)\right], \\
    \hat{\Sigma}_{ii} &= 1. \nonumber
\end{align}
The significance $Z$ evaluated on the grid becomes a multivariate Gaussian random vector $\hat{Z}$ with covariance $\hat{\Sigma}$:
\begin{align}
    \hat{Z} &\sim \mathcal{N}(\vec{0}, \hat{\Sigma}).
\end{align}
Similarly to the concept of the Asimov data set that may be used to estimate the expected local significance~\cite{cowan2011}, we propose to use a special set of background Asimov data sets to estimate the GP covariance, and subsequently estimate the global significance.

The \textit{set of Asimov data sets} is constructed as follows:
\begin{enumerate}
    \item Specify the binning of the data (for example invariant mass $m$) and the grid of scan points for the signal hypothesis (i.e\@. values for $M$).
    \item Construct the Asimov data set for the background model.
    \item For each chosen data bin produce a new data set, where a $1\sigma$ upward fluctuation is introduced in that bin of the background Asimov data set (see section~\ref{sec:examples} for concrete examples).
\end{enumerate}

We assume that the data in different bins are uncorrelated.\footnotemark[2]
To construct the approximate covariance, using eq.~(\ref{eq:sigcor_definition}),
we also use the fact that the covariance matrix of a sample of independent measurements can be decomposed
into a sum of the partial covariances, in our case calculated from the set of Asimov data sets:
\footnotetext[2]{%
There are situations when this assumption is not valid.
This happens, for example, in analyses that allow each event to be reconstructed
in multiple ways, and these reconstructed configurations contribute
to the distribution of the observable simultaneously.
One can also think of events recorded very closely in time becoming correlated due to out of time pile-up, when the detector
does not have time to relax to its initial state.}
\begin{align}
    &\hat{C}^a_{ij} = Z^a(M_i) \cdot Z^a(M_j), \\
    &\hat{C} = \sum_{a} \hat{C}^{a} \label{eq:cov_sum_rule},
\end{align}
where index $a$ enumerates Asimov data sets and $\hat{C}$ is a sum of partial covariances.
Note, however, the expression for the partial covariance $\hat{C}^a$ does not include the sample mean.
It was intentionally set to $0$, because we know the mean of the significance should be $0$.
We also know that variance of the significances is 1, so the final touch is to impose $1$ on the diagonal of $\hat{\Sigma}$ by rescaling the $\hat{C}$ as follows:
\begin{align}
    \hat{\Sigma}_{ij} &= \frac{\hat{C}_{ij}}{\sqrt{\hat{C}_{ii} \hat{C}_{jj}}}.
\end{align}
By normalizing this way, we again use the assumption that data in the bins are uncorrelated and we also assume that the covariance structure is approximately independent of the scale of a fluctuation.

\subsection{Number of upcrossings directly from the covariance matrix}\label{subsec:upcross_adhoc}
Given a differentiable kernel for a Gaussian Process, it is possible to calculate the average number of upcrossings
at any level without recourse to any random sampling~\cite{lutes2004,cramer1967}.
This represents a small advancement with respect to the Gross and Vitells extrapolation method.

To motivate the formula in eq.~(\ref{eq:direct_upcross_density}) for the density of upcrossings that we used directly in the form presented by Lutes et al.~\cite[Example~7.2]{lutes2004}, we first cite the generic expression for the density $\nu^{+}_{X}(u, t)$ of the average number of upcrossings of the level $u$ by any stochastic process $X(t)$~\cite[eq.~(7.3)]{lutes2004}:
\begin{align}\label{eq:upcross_density_generic}
    \nu^{+}_X(u, t) &= p_{X(t)}(u) \int_{0}^{\infty} \nu p_{\dot{X}(t)}[\nu \mid X(t) = u] \mathrm{d} \nu.
\end{align}
When $X(t)$ is a Gaussian process, $p_{X(t)}$ is the corresponding Gaussian density. The derivative $\dot{X}(t)$,
when conditioned on $X(t) = u$, follows a Gaussian distribution
with mean $\mu_\star$ and standard deviation $\sigma_\star$:
\begin{align}\label{eq:pxdot_params}
    &p_{\dot{X}(t)} [\nu \mid X(t) = u] = \mathcal{N}(\mu_\star(t), \sigma_\star(t))[\nu], \\
    &\mu_\star(t) = \frac{K_{\bullet}(t, t)}{K(t, t)}u = \rho_{\bullet}(t, t) \frac{\sigma_{\bullet}(t)}{\sigma(t)} u, \\
    &\sigma_\star(t) = \sigma_{\bullet} \sqrt{1 - \rho_{\bullet}^2(t, t)}, \nonumber
\end{align}
where $\sigma(t) = \sqrt{K(t, t)}$ is the standard deviation of the GP and $\mathcal{N}[\nu]$ is a Gaussian pdf evaluated at $\nu$.
The above expressions use a dot notation for derivatives along a single or both axes:
\begin{align}
    K_{\bullet}(x, y) &= \frac{\partial K(x, y)}{\partial y}, \;
    K_{\bullet \bullet}(x, y) = \frac{\partial^{2} K(x, y)}{\partial x\partial y}.
\end{align}
Consequently, the effective derivatives of the standard deviation and correlation are:
\begin{align}
    \sigma_{\bullet}(t) &= \sqrt{K_{\bullet \bullet}(t, t)}, \;
    \rho_{\bullet}(x, y) = \frac{K_{\bullet}(x, y)}{\sigma(x) \sigma_{\bullet}(y)}.
\end{align}

The expression for the density $\nu^{+}_X(u, t)$ of the average number of upcrossings at the level $u$ by the non-stationary GP with a covariance kernel $K$ and zero mean was derived by Lutes et al.~\cite[Example~7.2]{lutes2004} by integrating explicitly eq.~(\ref{eq:upcross_density_generic}) with the conditional Gaussian distribution $p_{\dot{X}}$ substituted from eq.~(\ref{eq:pxdot_params}) and is:
\begin{align}\label{eq:direct_upcross_density}
    &\nu^{+}_X(u, t) = \frac{\ee^{\frac{-u^2}{2 \sigma^{2}(t)}}}{\sqrt{2\pi} \sigma(t)} \left( \mu_{\star}(t) \Phi\left[\frac{\mu_{\star}(t)}{\sigma_{\star}(t)}\right] + \frac{\sigma_{\star}(t)}{\sqrt{2\pi}} \ee^{-\frac{\mu^{2}_{\star}(t)}{2 \sigma^{2}_{\star}(t)}}\right),
\end{align}
where $\Phi$ is the cumulative density function (CDF) of the standard normal distribution.

The analytic results above are valid for a continuous differentiable GP kernel. To evaluate these expressions numerically it proved useful to apply 2D spline interpolation to our covariance matrices before computing the derivatives. In this way we managed to reduce the uncertainty introduced by the coarse grid without sacrificing performance much, because spline interpolation provides exact spline derivatives straightaway. We then used Simpson quadrature to integrate the computed density over the mass range.

\subsection{Effect of coarse binning on upcrossings}\label{subsec:sneaks}
The Gross and Vitells upper bound is based on the average number of upcrossings of some level $u$ by the significance curve $Z(M)$.
When computing this number numerically, from MC toys, some finite resolution grid is used to approximate the curve.
It turns out that the resolution of the grid affects the resulting number of upcrossings.
In this section we investigate this effect, which is relevant for any Gross and Vitells procedure, whether it is based on brute force toys
or on the suggested GP toys.

\begin{figure}[ht]
    \centering
    \includegraphics[width=0.48\linewidth]{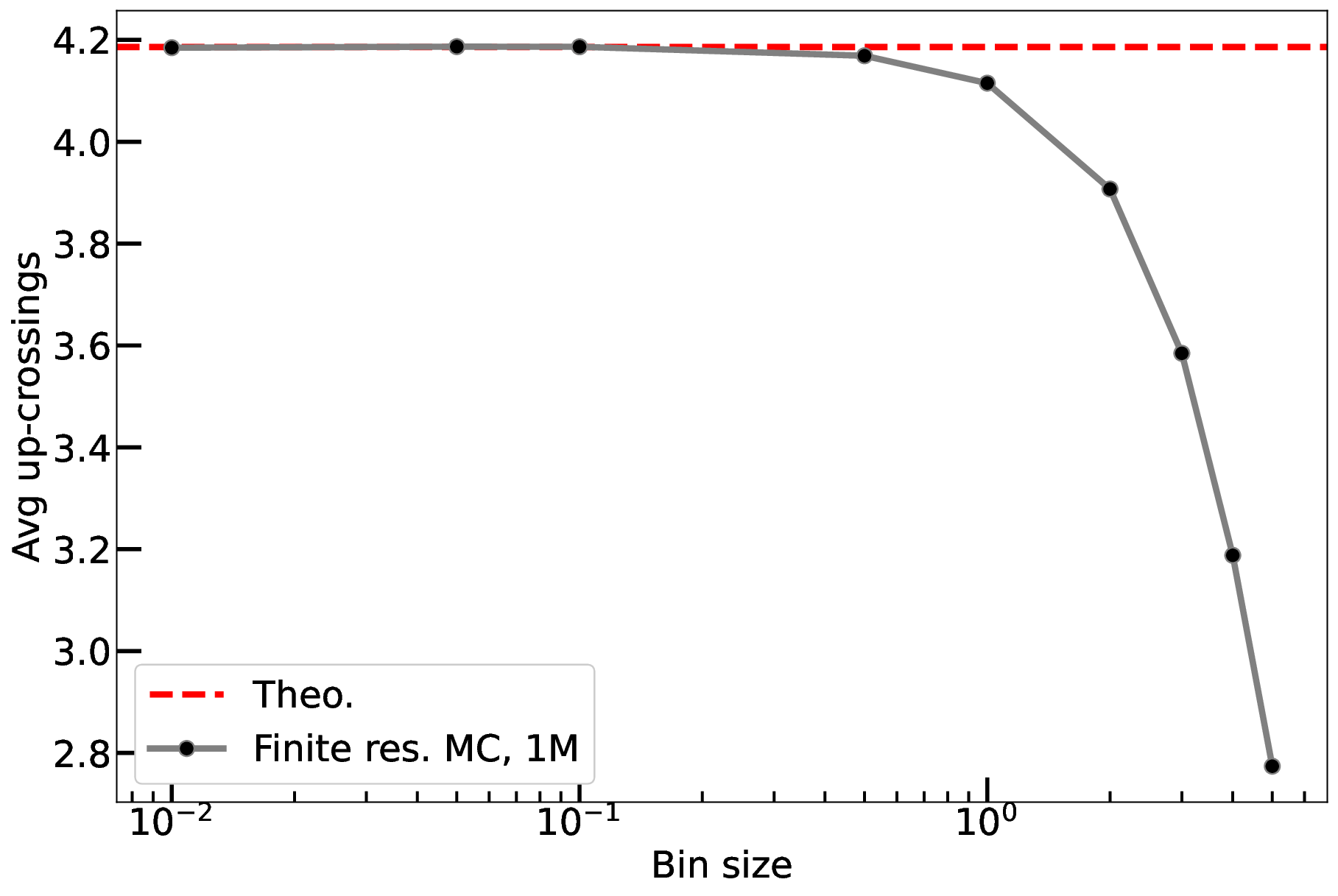}
    \caption{The average number of upcrossings as a function of bin size. The analytical estimation for the average number of upcrossings (red dashed line) was calculated following section~\ref{subsec:upcross_adhoc} for a squared exponential kernel (eq.~(\ref{eq:sneaks_kernel})). The average number of upcrossings for various bin sizes was estimated from $10^6$ GP toys (black dots connected by a gray line). The statistical errors are smaller than the size of the dots.}\label{fig:sneaks_upcross}
\end{figure}

We illustrate the effect on an example of an analytically defined Gaussian process with a squared exponential kernel:
\begin{align}\label{eq:sneaks_kernel}
    K(x, y) &= \ee^{-\frac{{(x-y)}^2}{\alpha^2}}, \quad \alpha^2 = 10.
\end{align}
We generated $10^6$ samples from this GP and counted the average number of upcrossings above the level $Z = 0.2$ on a set of grids with different resolutions. In figure~\ref{fig:sneaks_upcross} we show how the accuracy decreases with the resolution, starting when the bin size exceeds about one third of the correlation length.

The underestimation of the number of upcrossings affects the upper bound for the global significance, i.e.\@ the numerator of the trials factor, subsequently the trials factor itself will be underestimated. The local significance, however, which is the denominator, is also affected by the finite grid resolution. The tip of the peak at high significance most probably will appear between grid points, therefore the local significance will also be underestimated.

\begin{figure}[ht]
    \centering
    \includegraphics[width=0.48\linewidth]{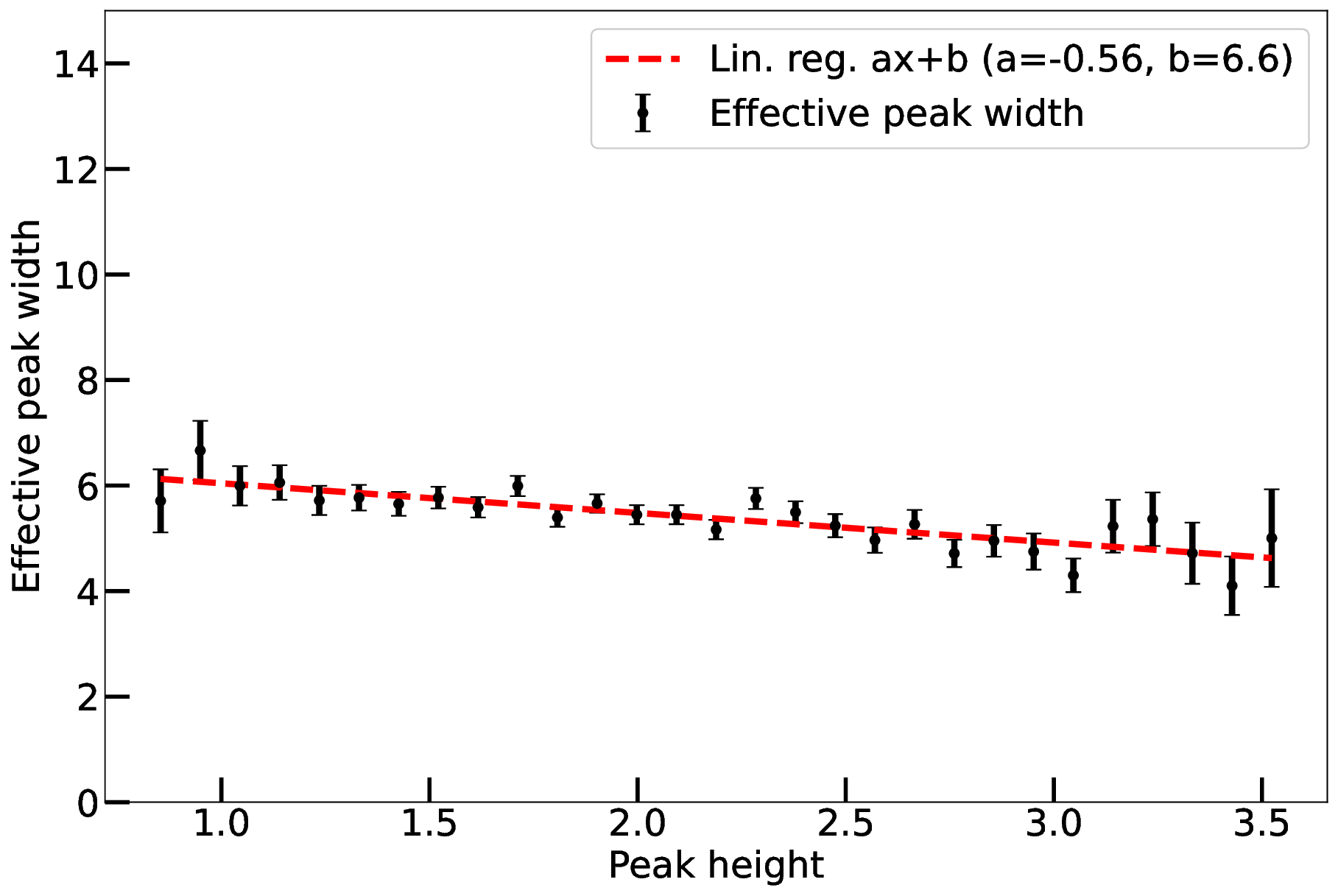}
\caption{The average effective width of the highest peak of GP samples for a squared exponential kernel (eq.~(\ref{eq:sneaks_kernel})) as a function of its height for $10^6$ GP toys  (black dots) fitted to a linear trend (red dashed line).}\label{fig:sneaks_peak_widths}
\end{figure}

We investigated the dependency of the effective peak width on its height. We sampled $10^6$ toys from the GP (eq.~(\ref{eq:sneaks_kernel})). For each sample we detected the highest point $(x_{max}, y_{max})$, chose a window of $40$ around it to select one peak and fitted it with a squared exponential shape $y = A^2 \ee^{{(x - B)}^2/C^2} + D$. For each sample we recorded a pair $(y_{max}, C)$, where $C$ is the effective peak width. We then removed the failed fit outliers by rejecting the $1\%$ lowest and $1\%$ highest heights. We observed that the average width of the peak decreases with its height (figure~\ref{fig:sneaks_peak_widths}), therefore, the finite grid effect on the trials factor gains a slight compensation.

With the Asimov set of samples we are not trying to improve this particular aspect of the Gross and Vitells approach, yet,
GP toys are also affected by the choice of the grid so we wanted to shed more light on this caveat.
Although we are not very concerned by the choice of a grid for the signal nuisance parameters,
of course it makes good sense to choose a grid that allows peaks to be described by at least a handful of points.

\section{Examples}\label{sec:examples}

The models we use as examples are inspired by searches for new resonances. The invariant masses of signal candidates are binned. Two of the search models have signals with specified widths whereas the third includes the signal width as a free parameter (on a grid) in the likelihood maximization.

\subsection{Background template model}\label{subsec:example_gv}

Gross and Vitells used this model in their study~\cite{gross2010}.
It consists of a fixed background shape modeled by the probability density function of the Rayleigh distribution with scale parameter $\sigma_b$, and a Gaussian signal with resolution $\sigma_s(M)$ that grows linearly with mass.
Background counts $D_i$ are sampled from the Poisson distribution with the rate $b_0\cdot b_i$ in each bin $i$ of the data grid $m_i$ defined as follows:\footnotemark[3]
\footnotetext[3]{$\sigma_b = 40 \, \GeV$, $b_0 = 2000$, $A = 2.5 \, \GeV, \; B = 50 \, \GeV$, $m_i = 0.5 - 154.5 \, \GeV$ with a step of $1\,\GeV$, $M_i = 5 - 120 \, \GeV$ with a step of $0.25 \, \GeV$.}
\begin{align}\label{eq:gv-bg_model}
    b_i &= N \frac{m_i}{\sigma_b} \ee^{-\frac{m_i^2}{2 \sigma_b^2}}, \\
    D_i &\sim \mathrm{Poisson}[b_0 \cdot b_i], \nonumber
\end{align}
where $N$ was chosen such that $\sum b_i = 1$ and, therefore, $b_0$ is the total number of events we expect to observe.

To test the precision of our GP-based approximation of the trials factor we compute the baseline via $10^6$ brute force MC simulations of the experiment. We scan each background simulation by testing signal hypotheses for $M$ in a range that is narrower, in this example, than the data grid $m$:
\begin{align}\label{eq:gv-s_model}
    s_i(M) &= \frac{1}{\sqrt{2 \pi} \sigma_s(M)} \ee^{-\frac{{(m_i - M)}^2}{2 \sigma_s^2{(M)}}}, \\
    \sigma_s(M) &= A (1 + \frac{M}{B}), \nonumber
\end{align}
and maximize the likelihood with Poisson statistics in the bins of data:
\begin{align}\label{eq:gv-sb_model}
    N_i &= \mu s_i(M) + \beta b_i, \\
    - \log \mathcal{L}(\mu, M, \beta) &= -\sum_i D_i \log(N_i) + \sum N_i, \nonumber
\end{align}
where $\mu$ and $\beta$ correspond to the numbers of inferred signal and background events that we vary to maximize the likelihood $\mathcal L$ with respect to data $D_i$, and $M$ is the nuisance parameter that defines the location of the hypothetical signal peak (the constant term of $\log\mathcal{L}$, which plays no role in the maximization, is not shown).

\begin{figure}[ht]
    \centering
    \begin{subfigure}[t]{0.48\linewidth}
        \includegraphics[width=\linewidth]{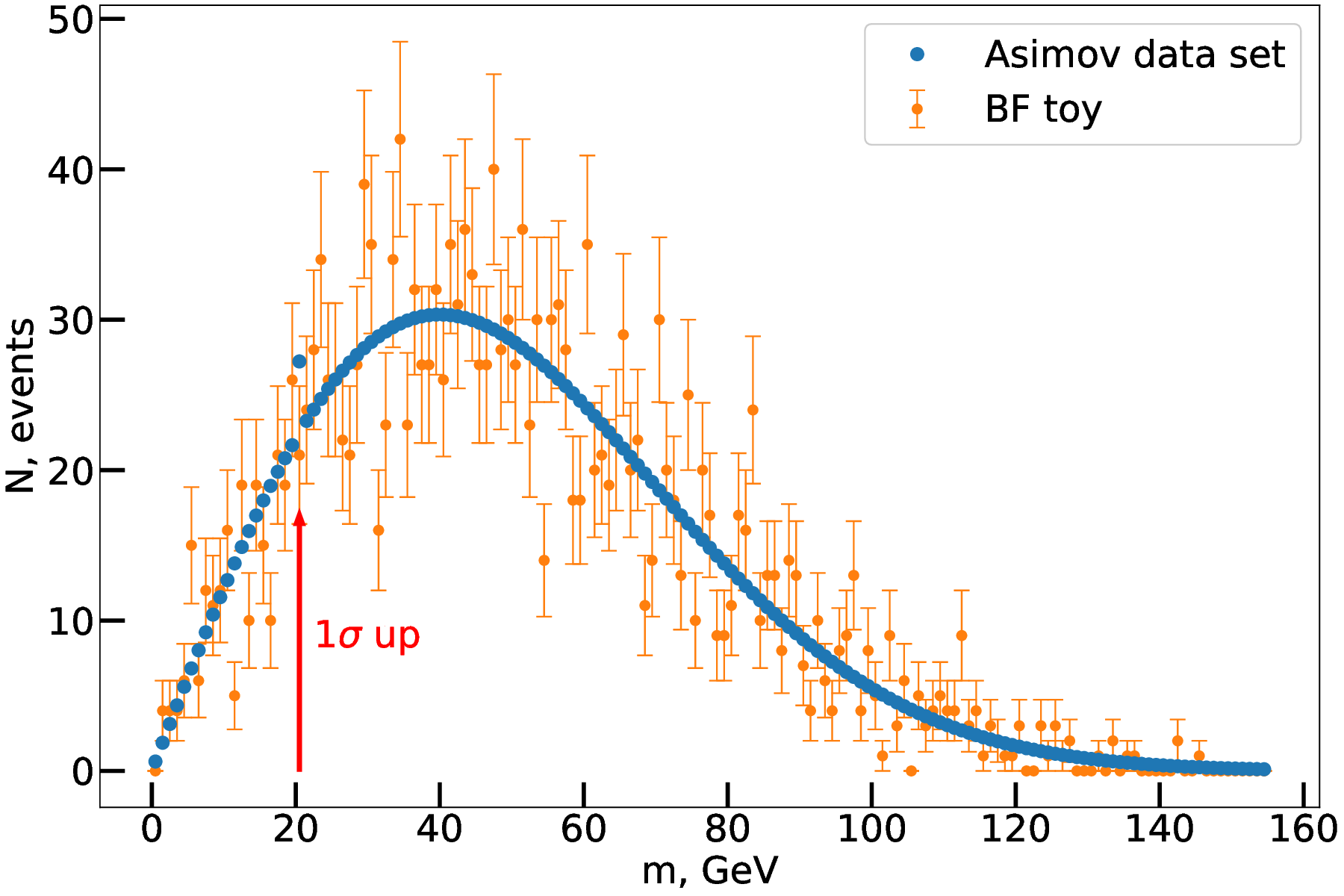}
        \caption{}\label{fig:gv-one_toy}
    \end{subfigure}%
    \begin{subfigure}[t]{0.48\linewidth}
        \includegraphics[width=\linewidth]{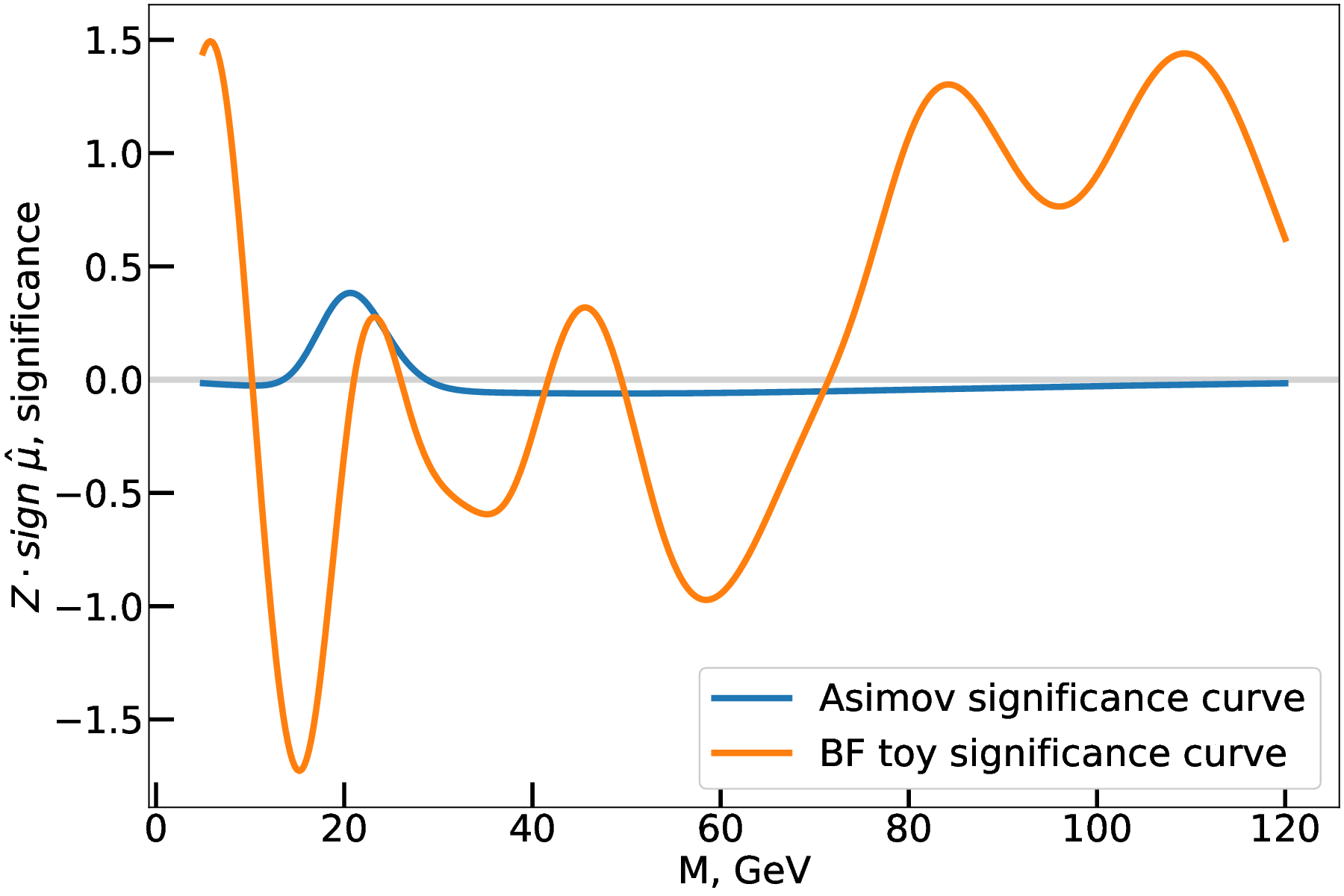}
        \caption{}\label{fig:gv-one_toy-significance}
    \end{subfigure}%
    \caption{A background sample, an Asimov toy and the corresponding significance curves from the Background template model. In (\subref{fig:gv-one_toy}) we show an example of a brute force toy sampled from the background template  (orange dots) and one data set from the set of Asimov data sets
    (blue dots). The red arrow specifies the bin that was distorted in this particular Asimov data set. In (\subref{fig:gv-one_toy-significance}) we show the significance curves from the signal plus background fits for the toy sample (orange line) and the Asimov data set (blue line).}\label{fig:gv-one_both}
\end{figure}

An example of a background simulation and the corresponding significance curve are shown in Fig~\ref{fig:gv-one_both}. As a cross-check we compute the average number of upcrossings for the significance threshold of $\sqrt{0.5}$, the same threshold Gross and Vitells used in their work. We get $4.3071 \pm 0.0016$ upcrossings, which is consistent with their result of $4.34 \pm 0.11$.

Next, we generate the set of Asimov background data sets. In this case there are $155$ of them, one for each bin of data. One of the Asimov data sets and the corresponding significance curve are shown in Fig~\ref{fig:gv-one_both} for comparison.

\begin{figure}[ht]
    \centering
    \begin{subfigure}[t]{0.48\linewidth}
        \includegraphics[width=\linewidth]{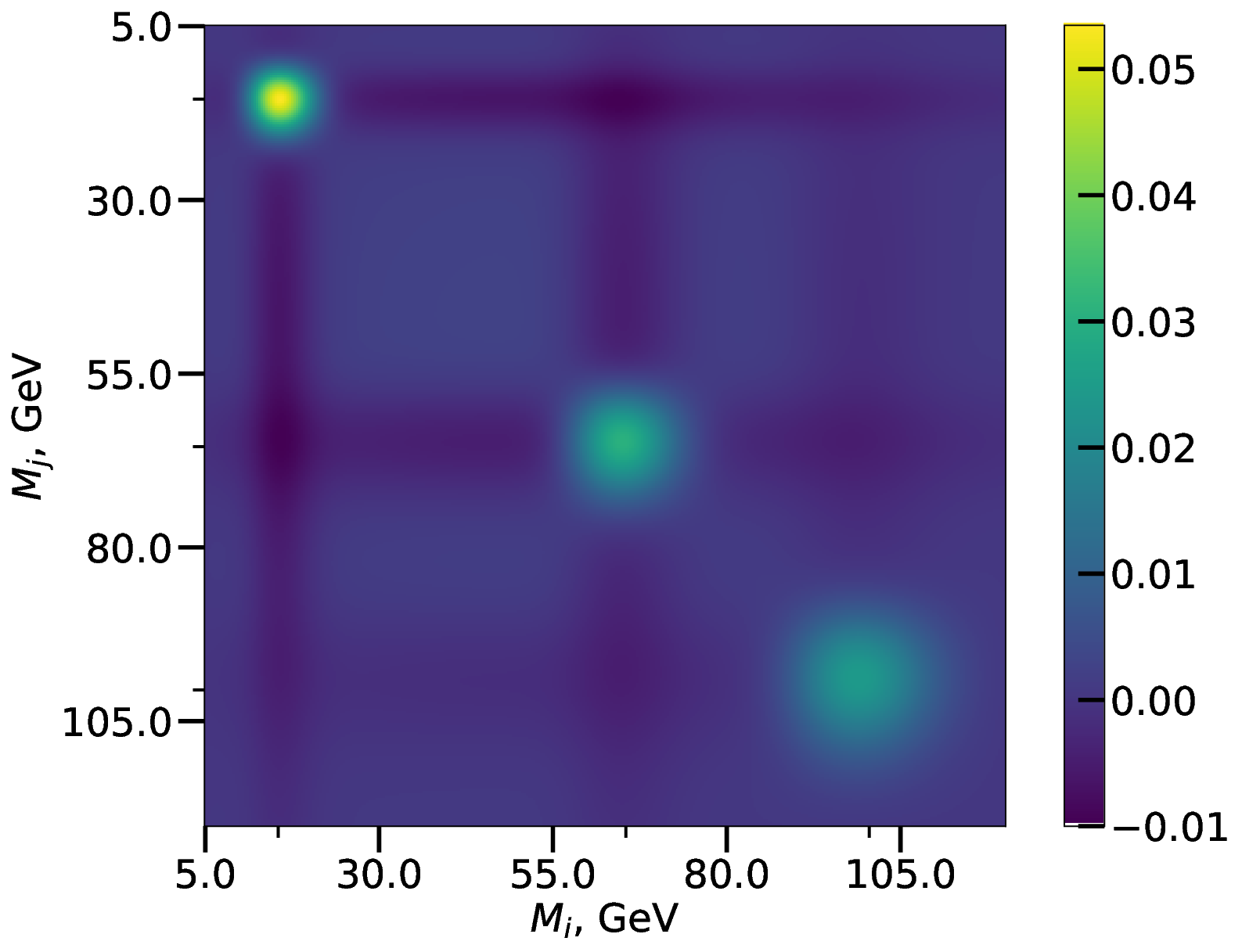}
        \caption{}\label{fig:gv-asimov_cov_partial}
    \end{subfigure}%
    \begin{subfigure}[t]{0.48\linewidth}
        \includegraphics[width=0.98\linewidth]{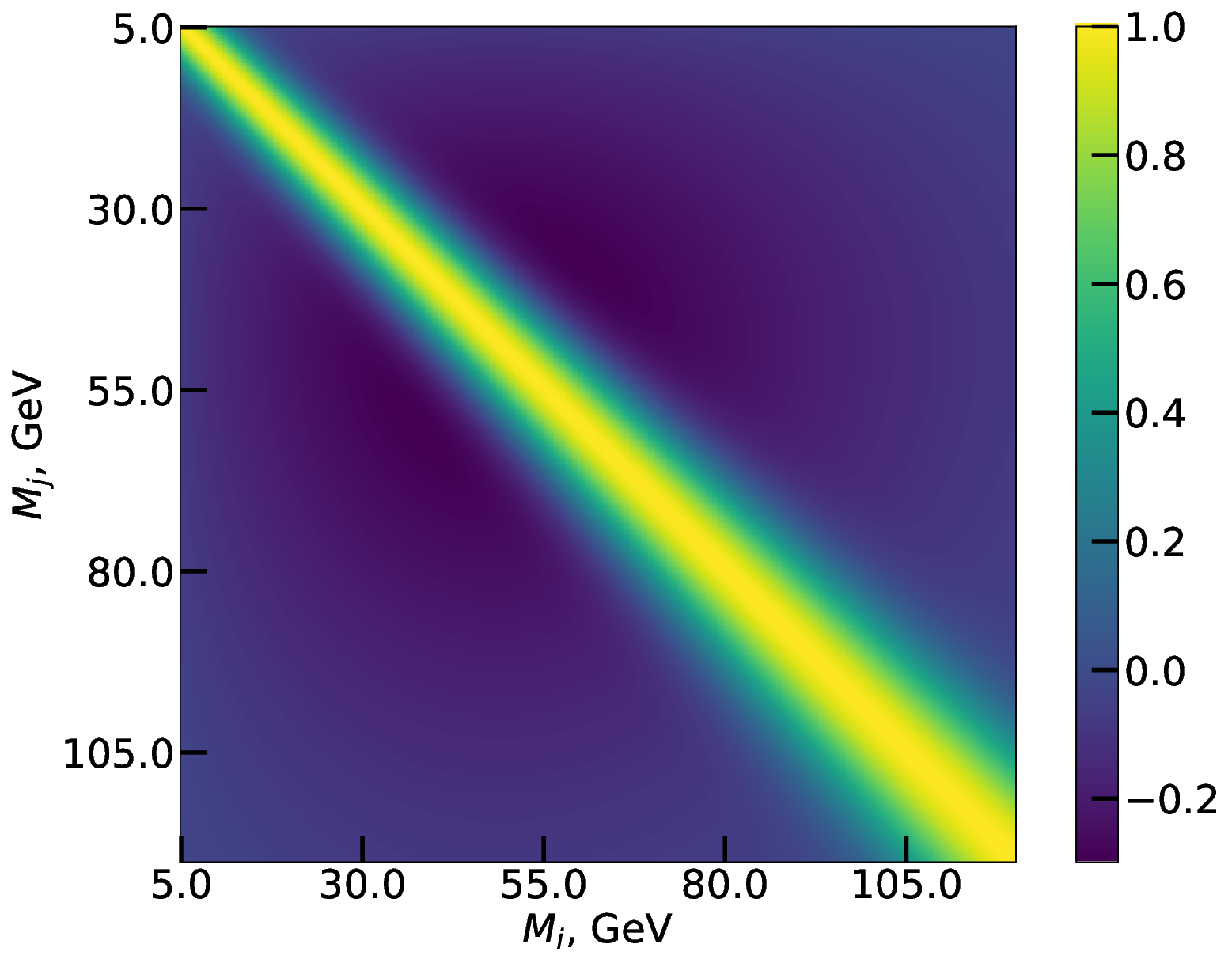}
        \caption{}\label{fig:gv-asimov_cov_full}
    \end{subfigure}
    \caption{Partial contributions to the covariance $\hat{\Sigma}(M_i, M_j)$ (\subref{fig:gv-asimov_cov_partial}) from $15.5$, $65.5$ and $100.5~\GeV$ Asimov data sets and the full Asimov GP covariance (\subref{fig:gv-asimov_cov_full}) for the Background template model.}
\end{figure}

For each Asimov data set we compute the partial covariance (see figure~\ref{fig:gv-asimov_cov_partial}), then we sum them and normalize according to eq.~(\ref{eq:cov_sum_rule}) to produce the Asimov GP covariance, shown in  figure~\ref{fig:gv-asimov_cov_full}. As expected, the width of the ridge on the diagonal of the covariance matrix is seen to have the same trend as the signal resolution that grows linearly with the mass hypothesis (eq.~(\ref{eq:gv-s_model})).

\begin{figure}[ht]
    \centering
    \begin{subfigure}[t]{0.48\linewidth}
        \includegraphics[width=0.96\linewidth]{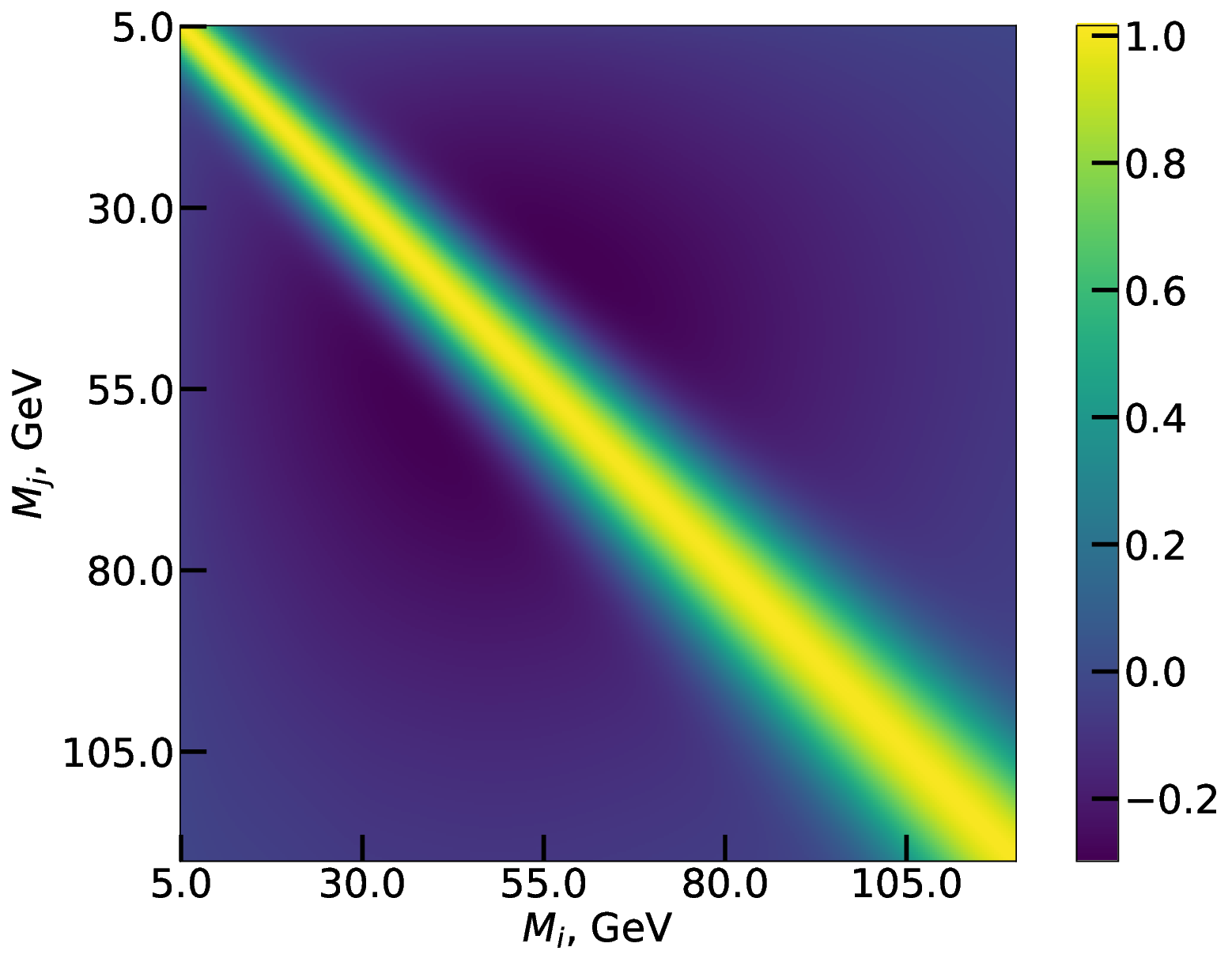}
        \caption{}\label{fig:gv-bf_cov_full}
    \end{subfigure}
    \begin{subfigure}[t]{0.48\linewidth}
        \includegraphics[width=\linewidth]{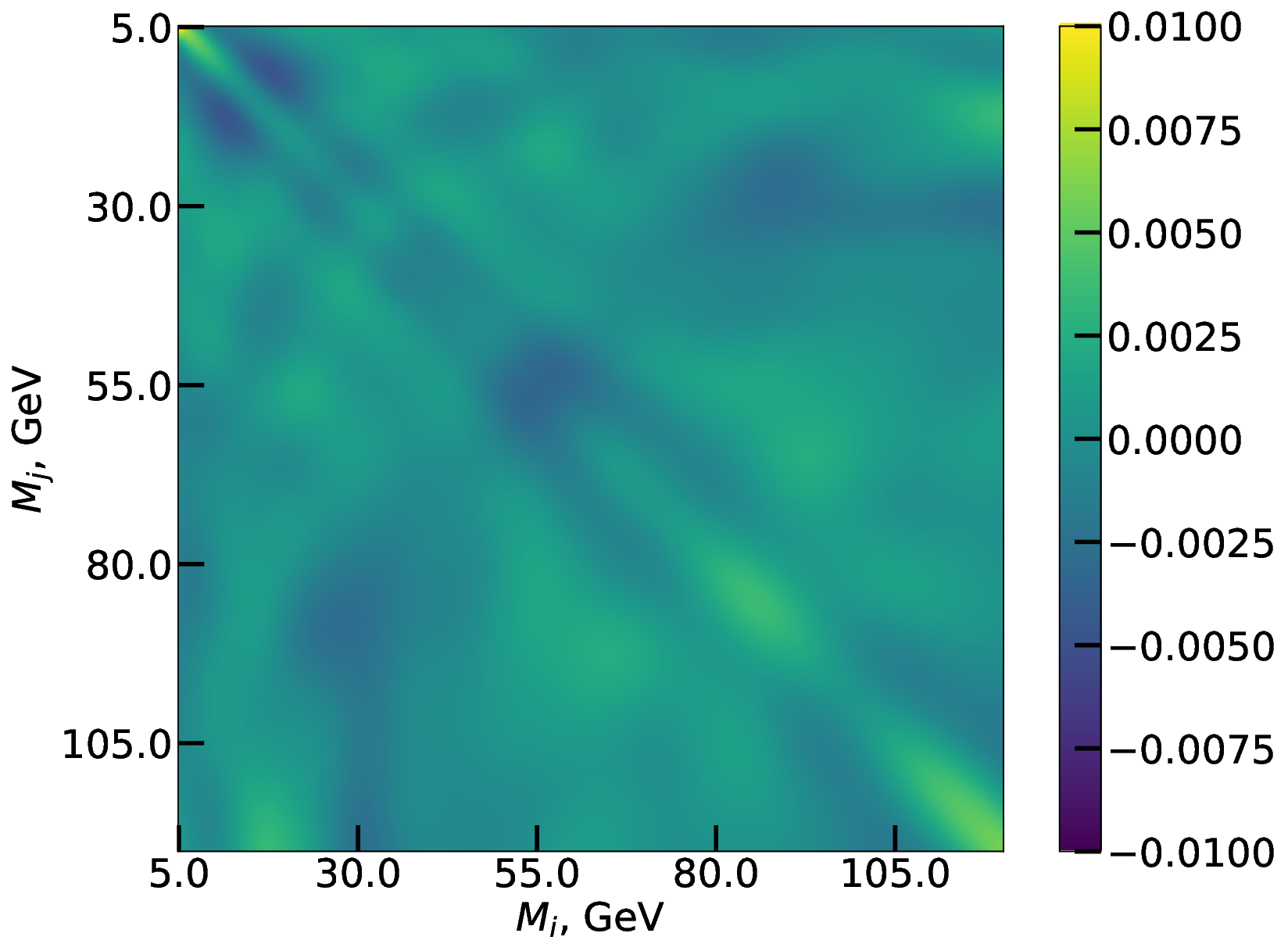}
        \caption{}\label{fig:gv-covcmp_diff}
    \end{subfigure}%
    \caption{The difference (\subref{fig:gv-covcmp_diff}) between the brute force (\subref{fig:gv-bf_cov_full}) and Asimov (figure~\ref{fig:gv-asimov_cov_full}) covariance matrices calculated for the Background template model.}
\end{figure}

Before using the covariance matrix to sample the significance curves, let us first evaluate how well we approximate the covariance matrix estimated from the brute force toys. For this we subtract the Asimov covariance, shown in  figure~\ref{fig:gv-asimov_cov_full}, from the covariance matrix constructed from $10^6$ brute force toys, shown in figure~\ref{fig:gv-bf_cov_full}, and plot the difference between the two covariances in figure~\ref{fig:gv-covcmp_diff}.
The differences between Asimov and brute force covariance are 100 times smaller than the values in the covariance itself.
We obtain similar patterns of differences for statistically independent simulations, so we conclude the pattern is systematic.

\begin{figure}[ht]
    \centering
    \includegraphics[width=0.48\linewidth]{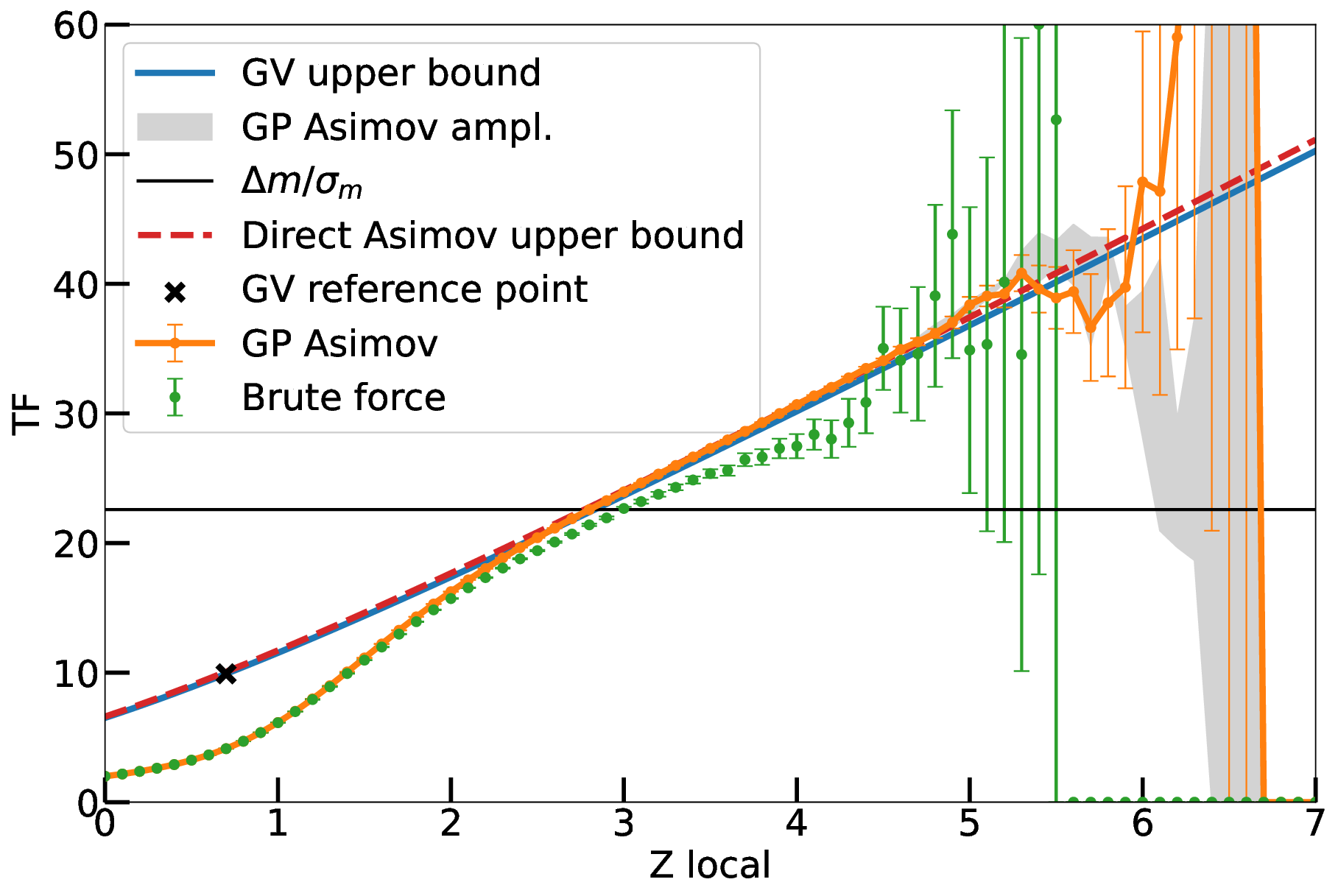}
    \caption{Trials Factor TF as a function of local significance $Z_\textrm{local}$ for the Background template model. We estimated TF from $10^6$ brute force toys (green dots) and compare it to the Asimov GP results (orange line). We color in gray the area enclosed between the two amplified lines from the TF sensitivity study for the Asimov covariance (section~\ref{subsec:example_gv}). The solid blue line shows the Gross and Vitells upper bound~\cite[eq.~(3)]{gross2010} extrapolated from the reference significance threshold $\sqrt{0.5}$ (marked by the black cross on the plot). For comparison we show the semi-analytical estimate (section~\ref{subsec:upcross_adhoc}) of the Gross and Vitells upper bound (red dashed line) and the rule of thumb (black line).}\label{fig:gv-trials_factor_cmp}
\end{figure}

In our experience, the trials factor appears to be the most sensitive metric to evaluate approximations of the global significance.
We estimate the trials factor from $360\cdot10^6$ significance curves sampled from the Asimov GP\@. In figure~\ref{fig:gv-trials_factor_cmp} we compare the Asimov GP trials factor, as a function of the local significance, to the brute force approach and approximations that include the Gross and Vitells upper bound and the rule of thumb estimate. The latter was calculated as a ratio of the mass range to the signal resolution, that we average, afterwards, over the mass range, since the resolution changes with the energy scale. We explore the sensitivity of the TF to the Asimov GP deviations by amplifying the difference between Asimov GP covariance and the brute force covariance by factors of $10$ upwards and downwards. From this simple sensitivity analysis we see that visible deviations only appear in the very high significance region (here $\sim6\sigma$), in which also the Asimov GP approach is affected by coarse binning effects, and in which the Gross and Vitells upper bound serves as a good approximation to the true TF value. In the same figure we also show that the analytic method described in section~\ref{subsec:upcross_adhoc} is in excellent agreement with the Gross and Vitells extrapolated upper bound.

We observe a significant deviation of the Asimov GP from the brute force trials factor estimate at high local significance levels. Qualitatively, we attribute this to the non-gaussianity of the errors introduced by Poissonian statistics that we approximated with a Gaussian distribution. Since the Poisson density is not symmetric around the mean, and is more pronounced for large upward fluctuations, the approximating Gaussian distribution gains upward bias. This effect is smaller at low local significances because the Poisson distribution can be well approximated by a quadratic shape around the mean. We confirmed this by studying the trials factors for larger integrated background rates and observing that the differences at high significance were reduced.

\subsection{\texorpdfstring{$H\rightarrow\gamma\gamma$}{H~→ ~γγ} inspired model}\label{subsec:example_hyy} This model is inspired by the ATLAS and CMS searches for the Higgs boson decaying to 2 photons~\cite{PhysRevLett.108.111803,CMSHyy}. In this case the background is modeled as an exponential distribution with Gaussian errors $\sigma_b$ in each bin of data. In order to challenge our method with a more realistic background scenario, in addition to the normalization $b_0$, the rate parameter $\alpha_b$ is also unconstrained in the maximum likelihood fits:\footnotemark[4]
\begin{align}\label{eq:hyy-bg_model}
    b_i(\alpha_b) &= \ee^{-(m_i - 100)\alpha_b}, \\
    D_i &\sim \mathcal{N}(b_0 \cdot b_i, \sigma_b). \nonumber
\end{align}
We also simplify the signal model by setting its resolution $\sigma_s$ to a constant:
\begin{align}
    s_i(M) &= \frac{1}{\sqrt{2 \pi} \sigma_s} \ee^{-\frac{{(m_i - M)}^2}{2 \sigma_s^2}}.
\end{align}
We unify the grids and use the same bins for both sampling of the MC toys ($m_i$) and signal location scan ($M_i$). The structure of the likelihood function in this case is:
\footnotetext[4]{$\alpha_b = 0.033 \, \GeV^{-1}$, $b_0 = 10$, $\sigma_b = 0.3$, $\sigma_s = 5 \, \GeV$, $m_i = M_i = 100 - 160 \, \GeV$ with a step of $1 \, \GeV$.}
\begin{align}\label{eq:hyy-sb_model}
    N_i &= \mu s_i(M) + \beta b_i(\alpha_b), \\
    - \log \mathcal{L}(\mu, M, \beta, \alpha_b) &= \sum_i {(N_i - D_i)}^2. \nonumber
\end{align}

\begin{figure}[ht]
    \centering
    \begin{subfigure}[t]{0.48\linewidth}
        \includegraphics[width=\linewidth]{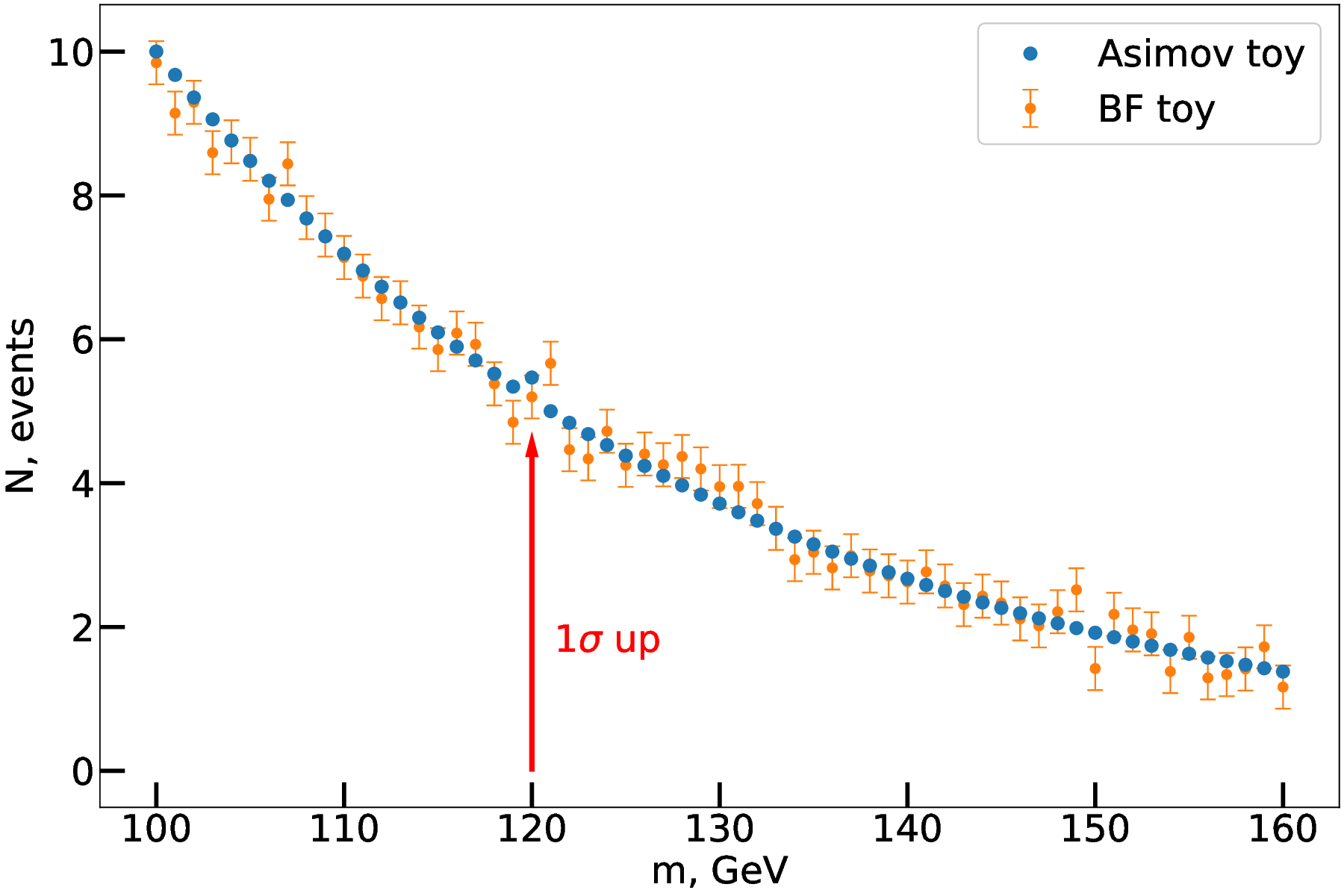}
        \caption{}\label{fig:hyy-one_toy}
    \end{subfigure}%
    \begin{subfigure}[t]{0.48\linewidth}
        \includegraphics[width=\linewidth]{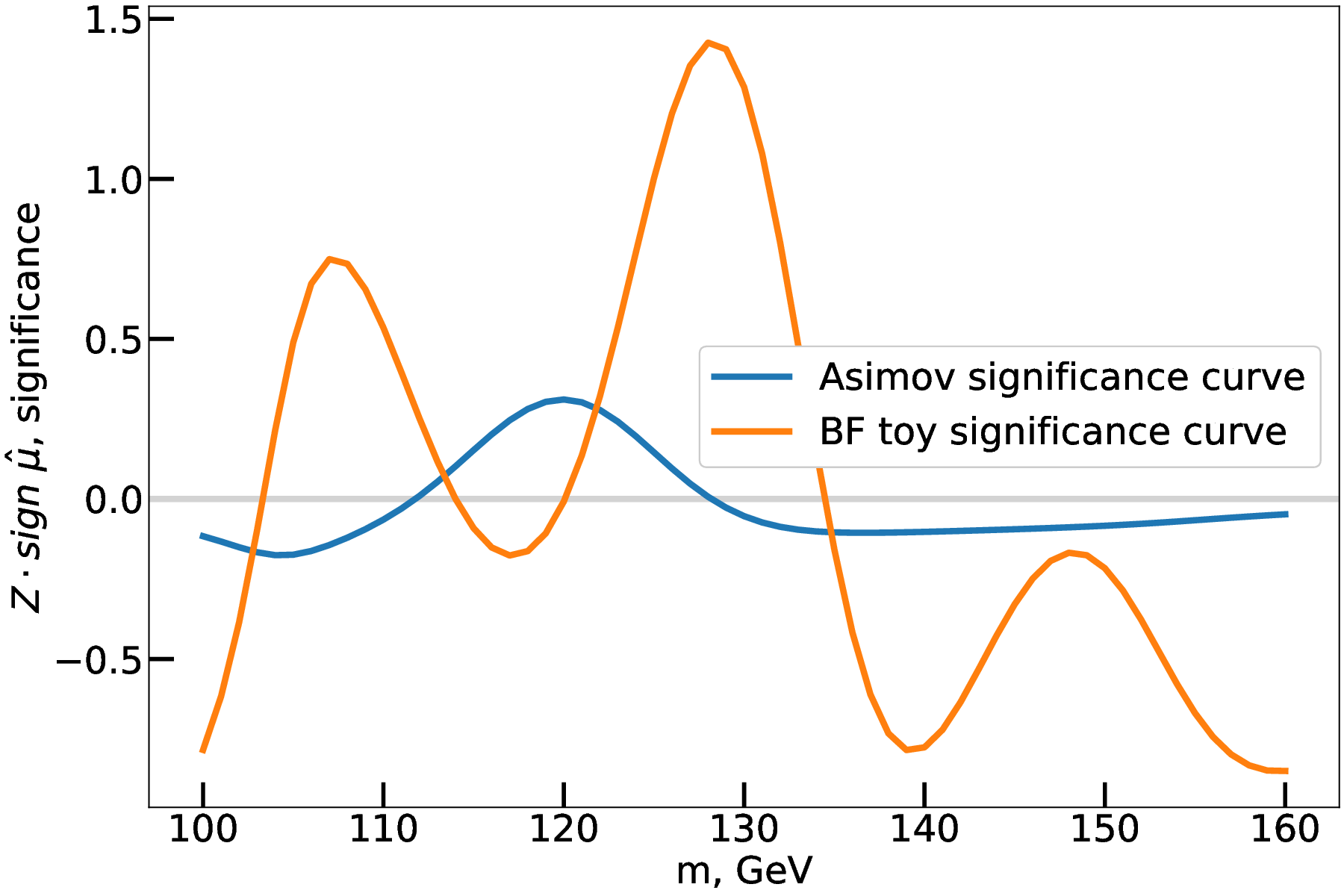}
        \caption{}\label{fig:hyy-one_toy-significance}
    \end{subfigure}%
    \caption{A background sample, an Asimov toy (\subref{fig:hyy-one_toy}) and the corresponding significance curves (\subref{fig:hyy-one_toy-significance}) from the $H\rightarrow\gamma\gamma$ inspired model. See the detailed explanation of the two plots in figure~\ref{fig:gv-one_both}.}
\end{figure}

Following the same algorithm as for the Gross and Vitells model, we first show an example of the brute force and Asimov background samples (figure~\ref{fig:hyy-one_toy}) together with the corresponding significance curves (figure~\ref{fig:hyy-one_toy-significance}).

\begin{figure}[ht]
    \centering
    \begin{subfigure}[t]{0.48\linewidth}
        \includegraphics[width=\linewidth]{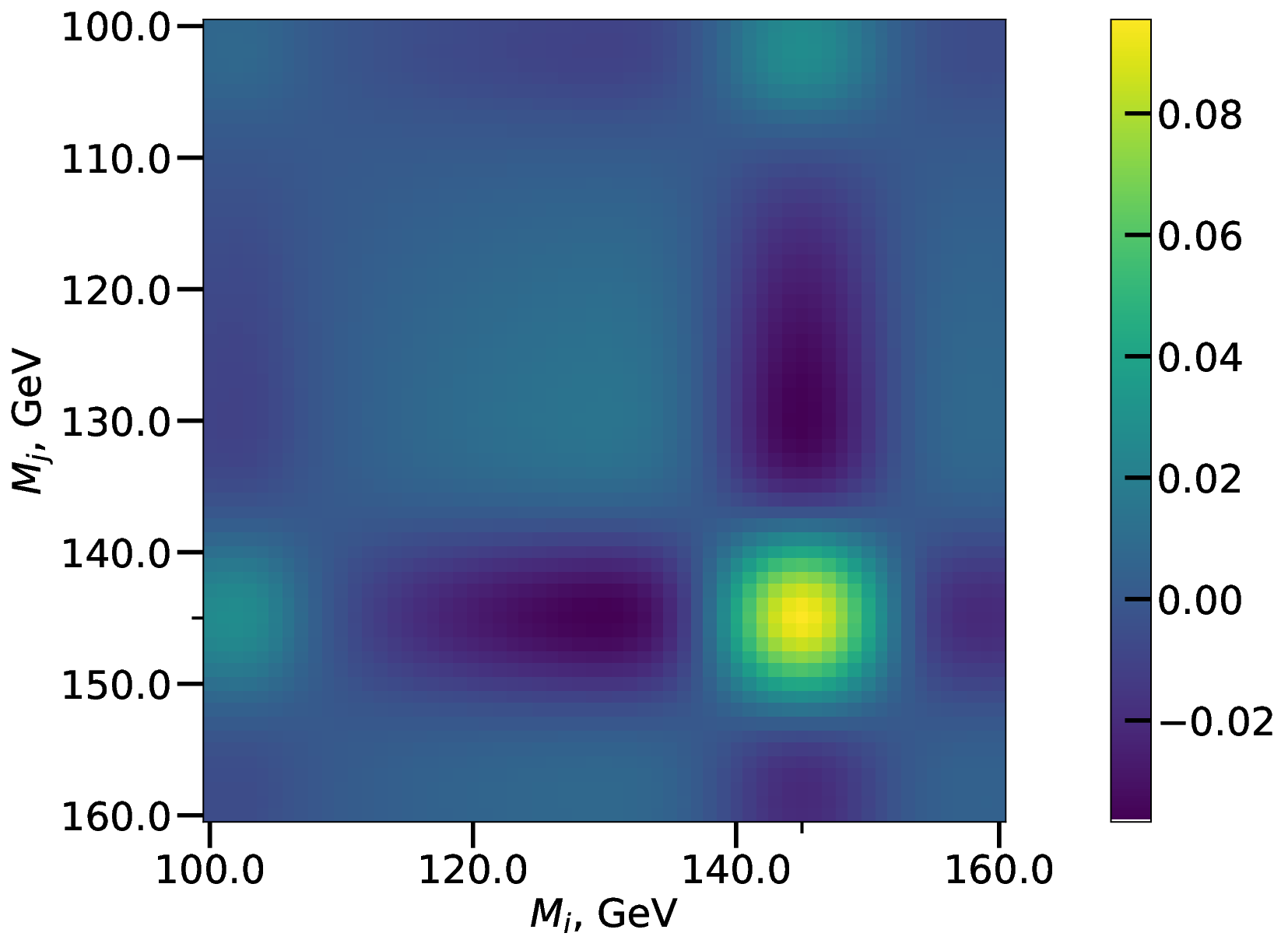}
        \caption{}\label{fig:hyy-asimov_cov_partial}
    \end{subfigure}%
    \begin{subfigure}[t]{0.48\linewidth}
        \includegraphics[width=\linewidth]{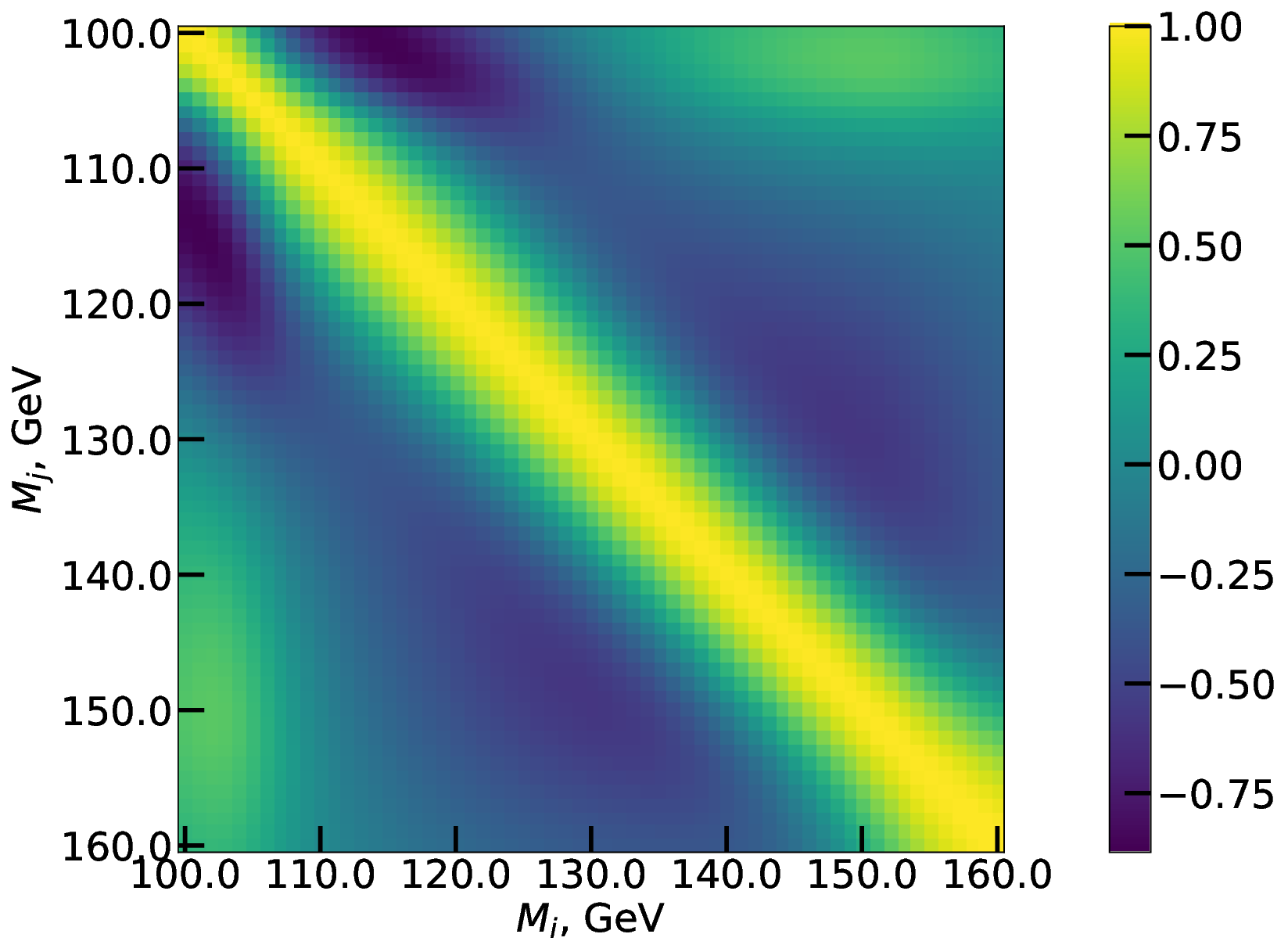}
        \caption{}\label{fig:hyy-asimov_cov_full}
    \end{subfigure}%
    \caption{The partial contribution to the covariance $\hat{\Sigma}(M_i, M_j)$ (\subref{fig:hyy-asimov_cov_partial}) from the $145.5~\GeV$ Asimov data set and the full Asimov GP covariance (\subref{fig:hyy-asimov_cov_full}) for the $H\rightarrow\gamma\gamma$ inspired model.}
\end{figure}

Already from the background fit to the Asimov sample (figure~\ref{fig:hyy-one_toy-significance}), we can see that the increased flexibility of the background model introduces more sophisticated long-range correlations. Notice the asymmetry between the left and right sides of the peak. When the signal hypothesis is near the Asimov fluctuation at $120~\GeV$, the signal part of the model accommodates it, but well below and above this the rate parameter of the background model tries to compensate for the local excess, which distorts the background model and leads to weak but clear regions of long-range anti-correlation and correlation. We confirm this observation from the plots of the GP covariance that we estimate from the Asimov samples (figure~\ref{fig:hyy-asimov_cov_full}).

\begin{figure}[ht]
    \centering
    \begin{subfigure}[t]{0.48\linewidth}
        \includegraphics[width=0.98\linewidth]{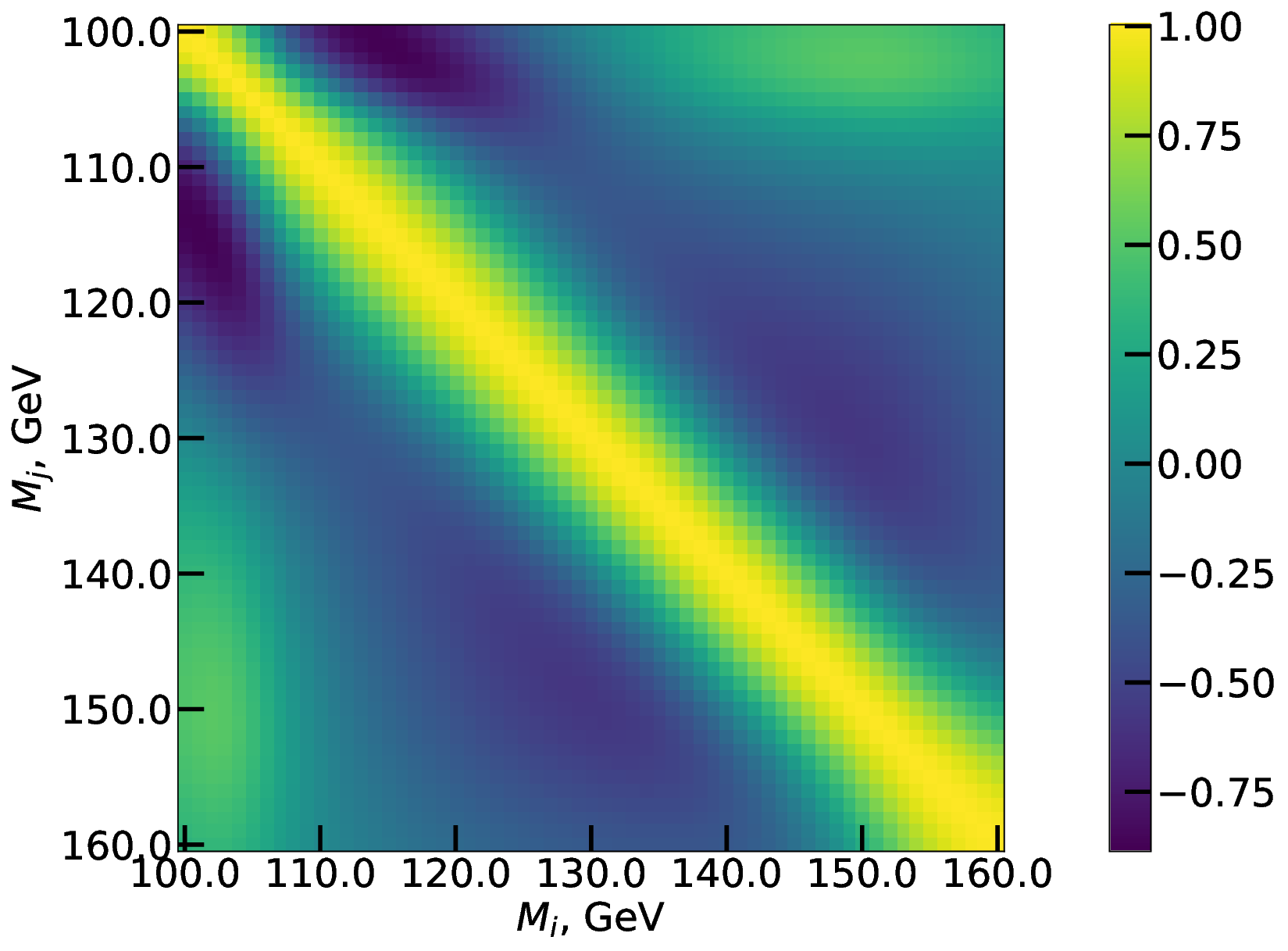}
        \caption{}\label{fig:hyy-bf_cov_full}
    \end{subfigure}%
    \begin{subfigure}[t]{0.48\linewidth}
        \includegraphics[width=\linewidth]{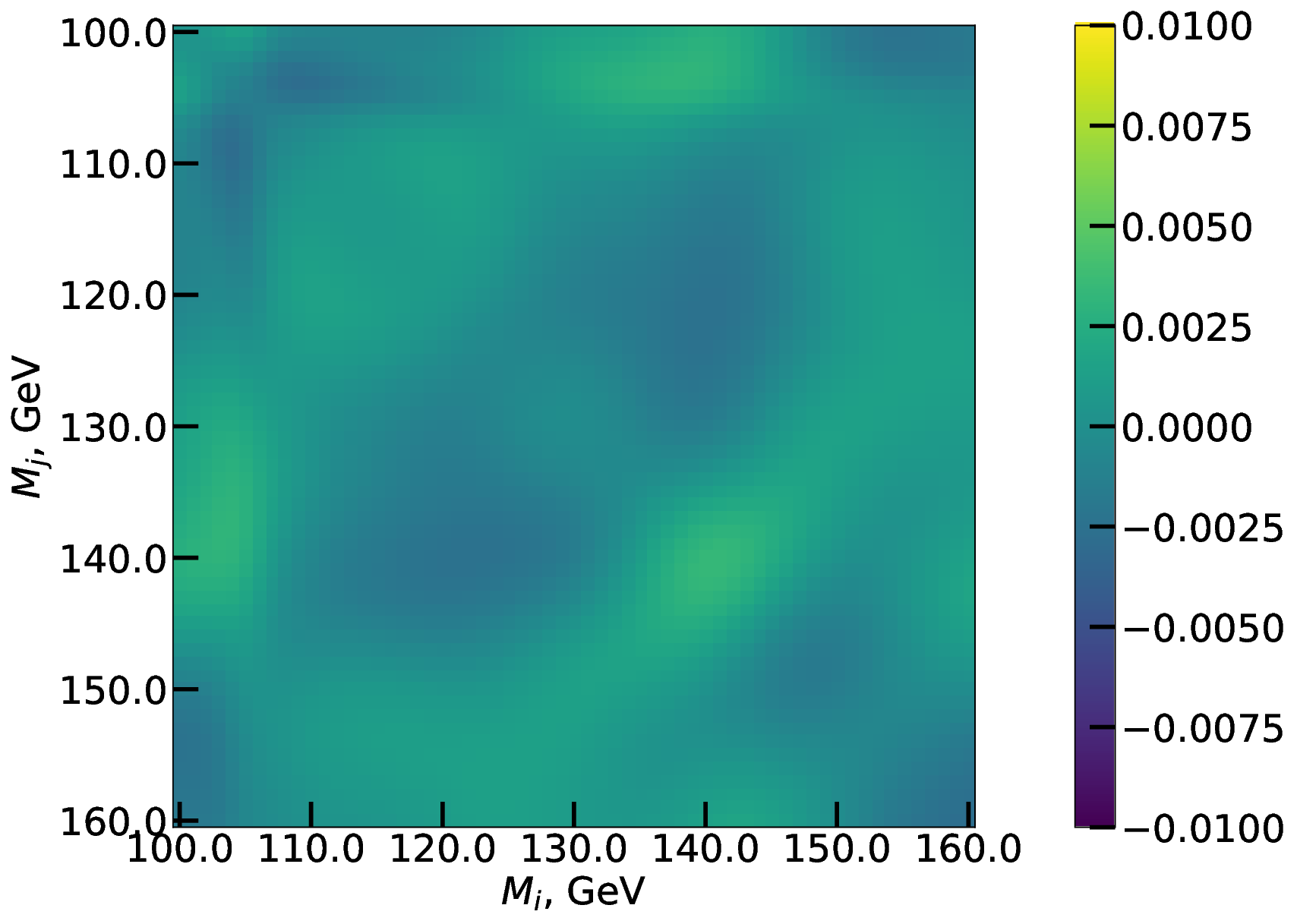}
        \caption{}\label{fig:hyy-covcmp_diff}
    \end{subfigure}%
    \caption{The difference (\subref{fig:hyy-covcmp_diff}) between the brute force (\subref{fig:hyy-bf_cov_full}) and Asimov (figure~\ref{fig:hyy-asimov_cov_full}) covariance matrices calculated for the $H\rightarrow\gamma\gamma$ inspired model.}\label{fig:hyy-covcmp}
\end{figure}

We then compare the Asimov covariance to the one calculated with brute force (figure~\ref{fig:hyy-covcmp}). It shows the same degree of accuracy as we have observed for the Background template model: the differences between the Asimov and brute force covariances are 100 times smaller than the values of the covariance itself.

\begin{figure}[ht]
    \centering
    \includegraphics[width=0.48\linewidth]{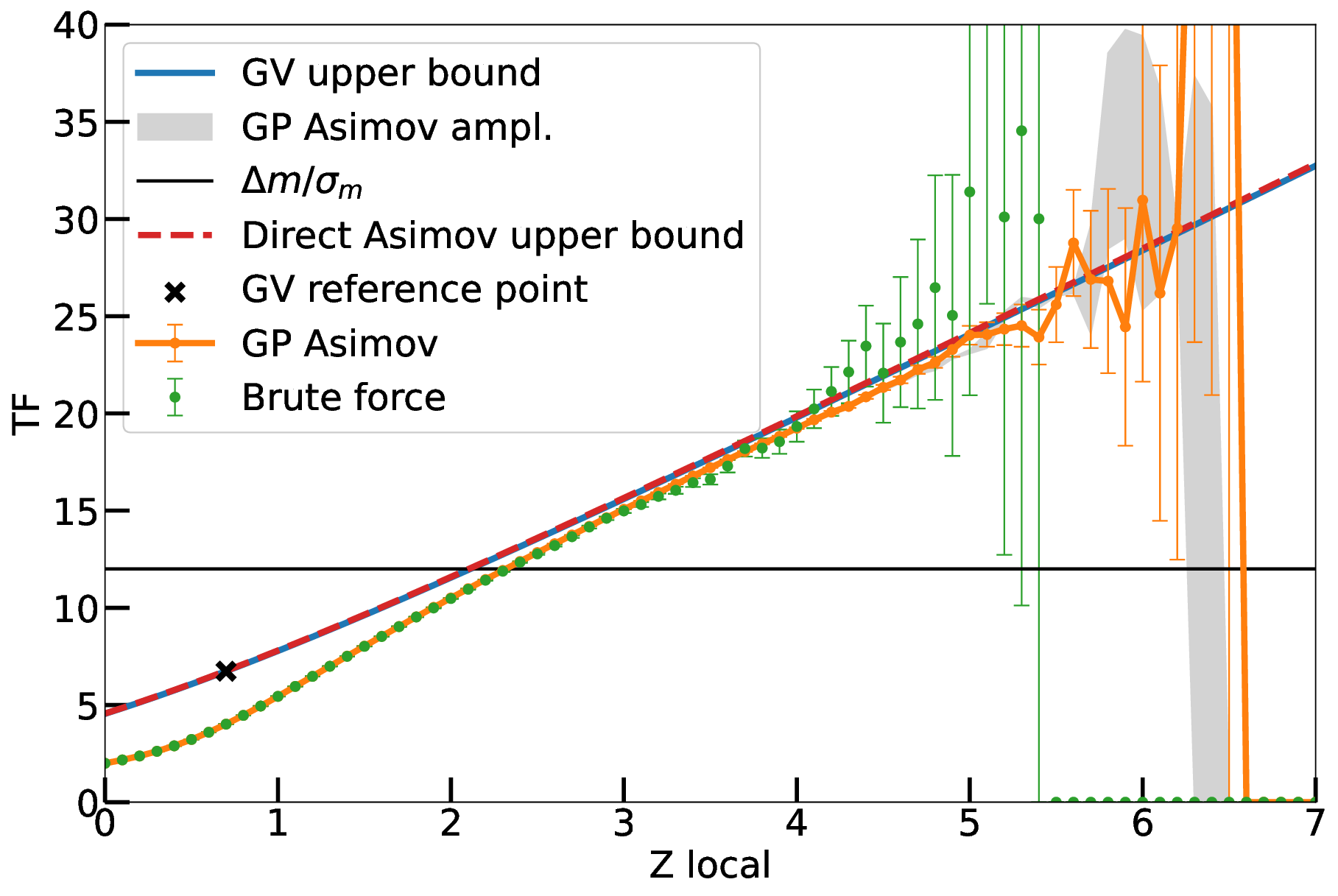}
    \caption{Trials Factor TF as a function of local significance $Z_\textrm{local}$ for the $H\rightarrow\gamma\gamma$ inspired model. See the detailed explanation of the plot in figure~\ref{fig:gv-trials_factor_cmp}.}\label{fig:hyy-trials_factor_cmp}
\end{figure}

The various computations of the trials factor for the $H\rightarrow\gamma\gamma$ inspired model are shown in figure~\ref{fig:hyy-trials_factor_cmp}. Since the likelihood and the generation of Monte Carlo data sets both use a Gaussian pdf for the data in each bin, there is no significant difference between the brute force and GP results at high significance, in contrast to the results for the Background template model.

\subsection{\texorpdfstring{$H\rightarrow\gamma\gamma$}{H~→ ~γγ} model with 2-dimensional search region}\label{subsec:example_hyy2d}
We keep the background model unchanged from the 1-D example (eq.~(\ref{eq:hyy-bg_model})), however, we want to study a signal model with more than one nuisance parameter. In this case a 1-dimensional significance curve becomes a multidimensional surface. However, each point on this surface still behaves like a standard normal random variable. As defined in section~\ref{sec:method}, $M$ denotes the set of points that constitute the search region.

For example, in the 2-D case, when in addition to \textit{mass} $m_s$ we also add \textit{width} $\sigma_s$ as a second nuisance parameter, $M$ gains a second axis, consequently, the significance surface becomes 2-dimensional. The signal model for this example is\footnotemark[5]
\footnotetext[5]{$\alpha_b = 0.033 \, \GeV$, $b_0 = 10$, $\sigma_b = 0.3 \, \GeV$, $M_{ij} = (a_i, b_j)$, $m_i = 100 - 160 \, \GeV$ with a step of $1 \, \GeV$; $a_i = 100 - 160 \, \GeV$ and $b_i = 1-10 \, \GeV$ with 61 points on the grid each.}
\begin{align}
    s_i(M) = s_i(m_s, \sigma_s) &= \frac{1}{\sqrt{2 \pi} \sigma_s} \ee^{-\frac{{(m_i - m_s)}^2}{2 \sigma_s^2}}.
\end{align}
The likelihood function is the same as for the 1-D case (see eq.~(\ref{eq:hyy-sb_model})).

\begin{figure}[ht]
    \centering
    \begin{subfigure}[t]{0.36\linewidth}
        \includegraphics[width=\linewidth]{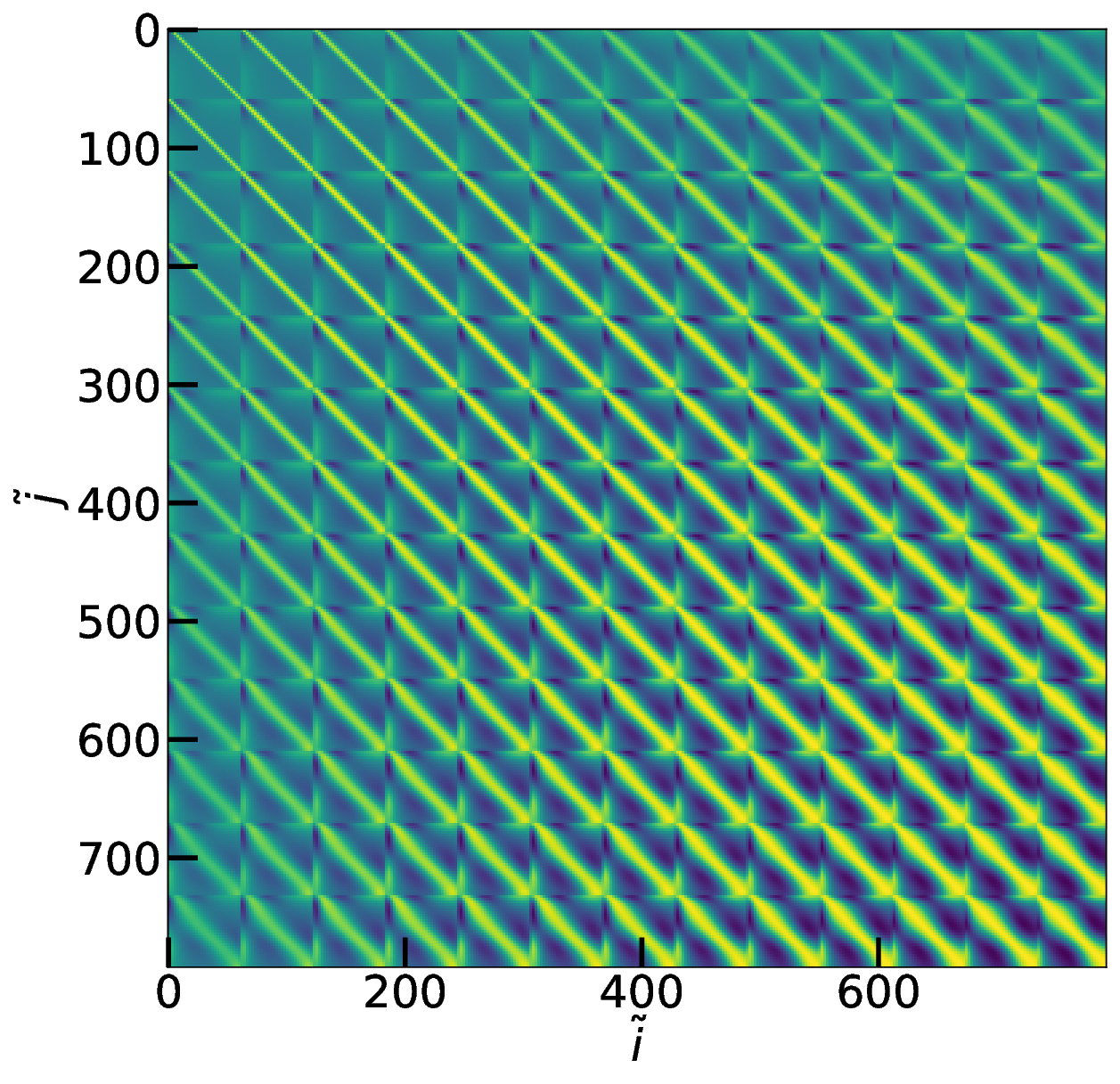}
        \caption{}\label{fig:hyy2d-bf_cov_full-mass}
    \end{subfigure}\hspace{0.05\linewidth}%
    \begin{subfigure}[t]{0.36\linewidth}
        \includegraphics[width=\linewidth]{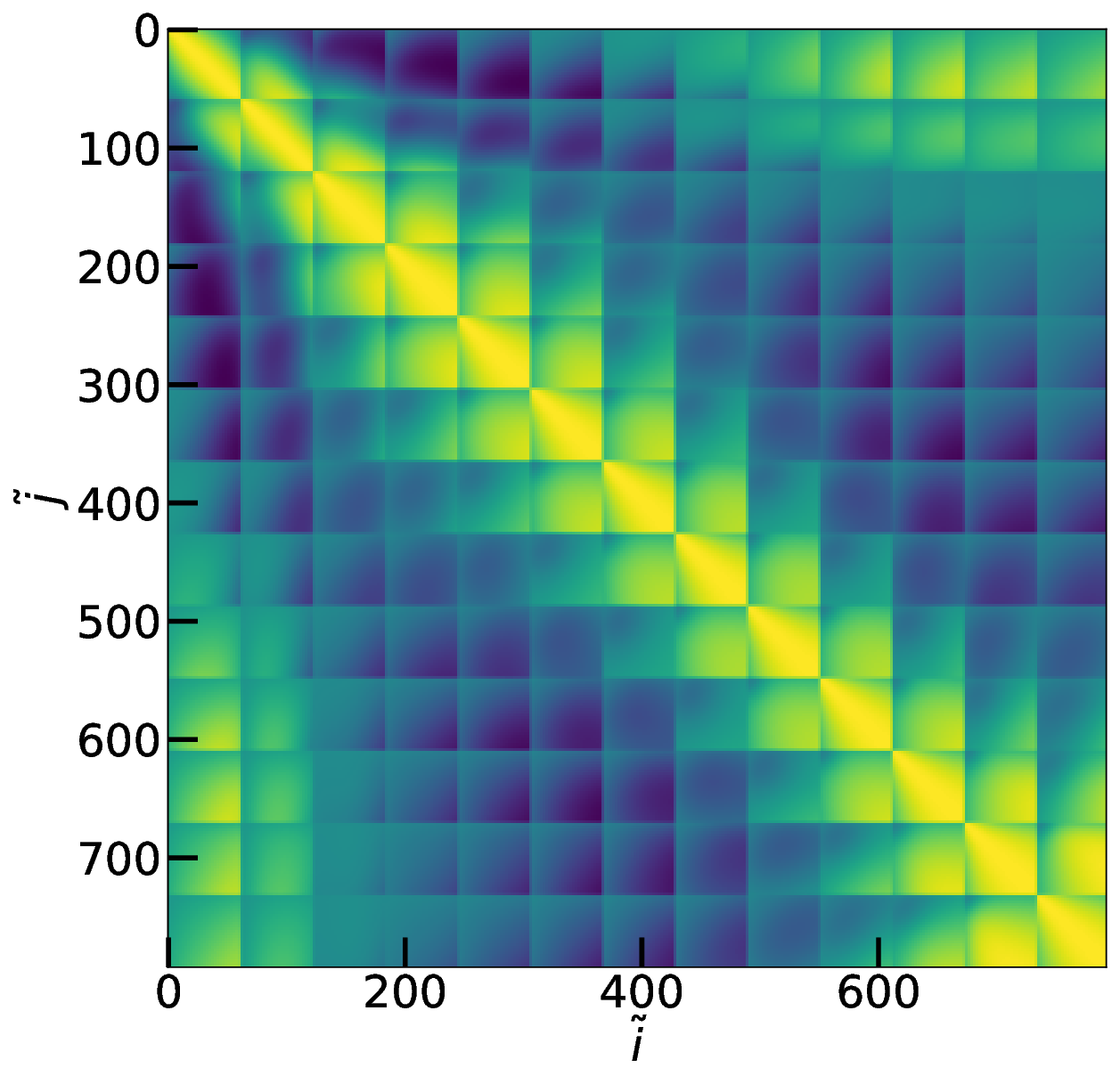}
        \caption{}\label{fig:hyy2d-bf_cov_full-width}
    \end{subfigure}%
    \caption{4D covariance matrix $M_{\tilde{i}\tilde{j}}$ unwrapped into 2D following eq.~(\ref{eq:2dcov_unwrap}) (\subref{fig:hyy2d-bf_cov_full-mass}) and its transposed version (\subref{fig:hyy2d-bf_cov_full-width}), for the $H\rightarrow\gamma\gamma$ inspired model with a 2D search region. For the purpose of illustration, the less frequently changing dimension on each plot has a reduced number of points, consequently, the plots have a reduced number of ``squares''. The notation $M_{\tilde{i}\tilde{j}}$ emphasises that the grid was artificially downsampled.}\label{fig:hyy2d-cov_full}
\end{figure}

\begin{figure}[ht]
    \centering
    \begin{subfigure}[t]{0.45\linewidth}
        \includegraphics[width=\linewidth]{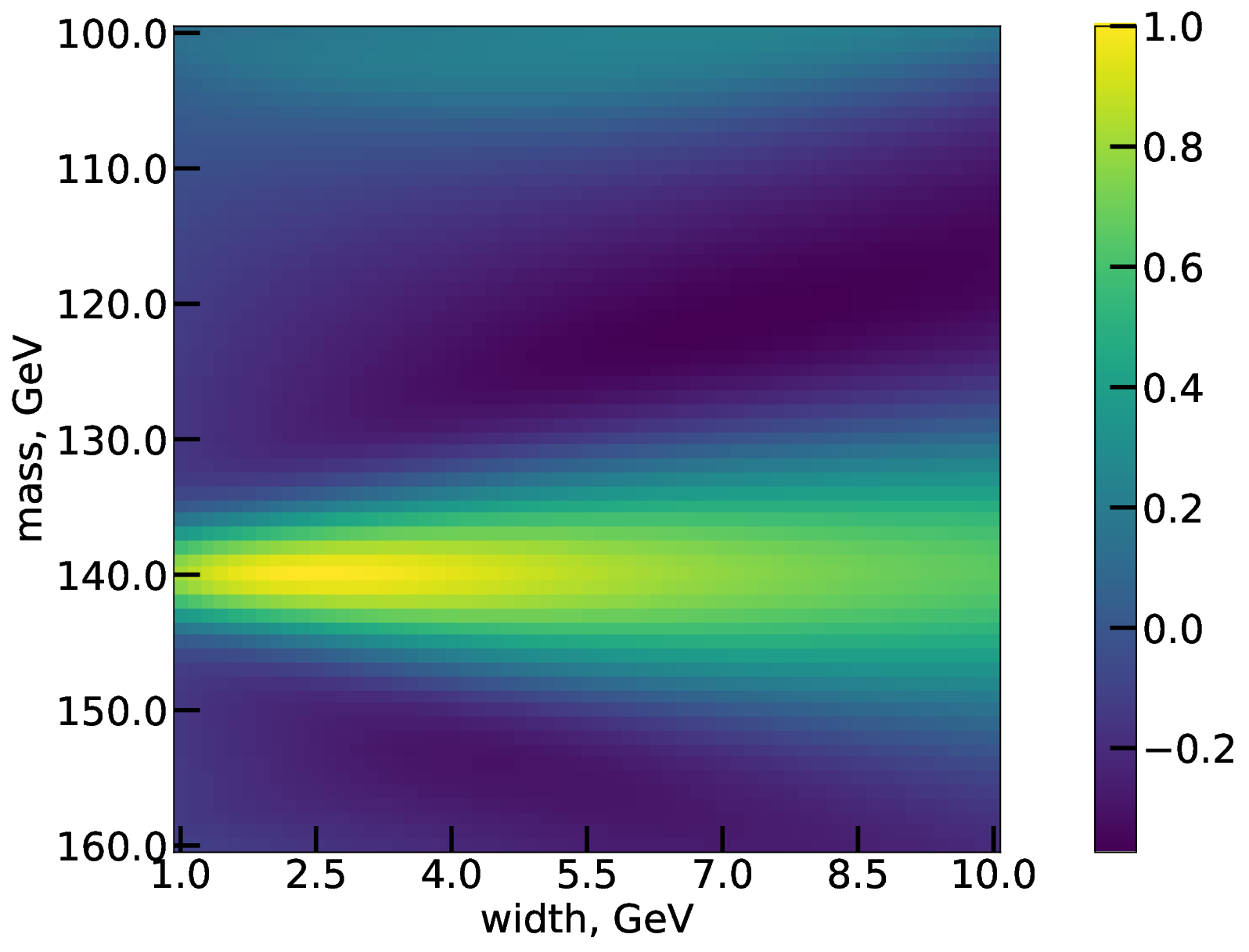}
        \caption{}\label{fig:hyy2d-covcmp_mass-width}
    \end{subfigure}%
    \begin{subfigure}[t]{0.454\linewidth}
        \includegraphics[width=\linewidth]{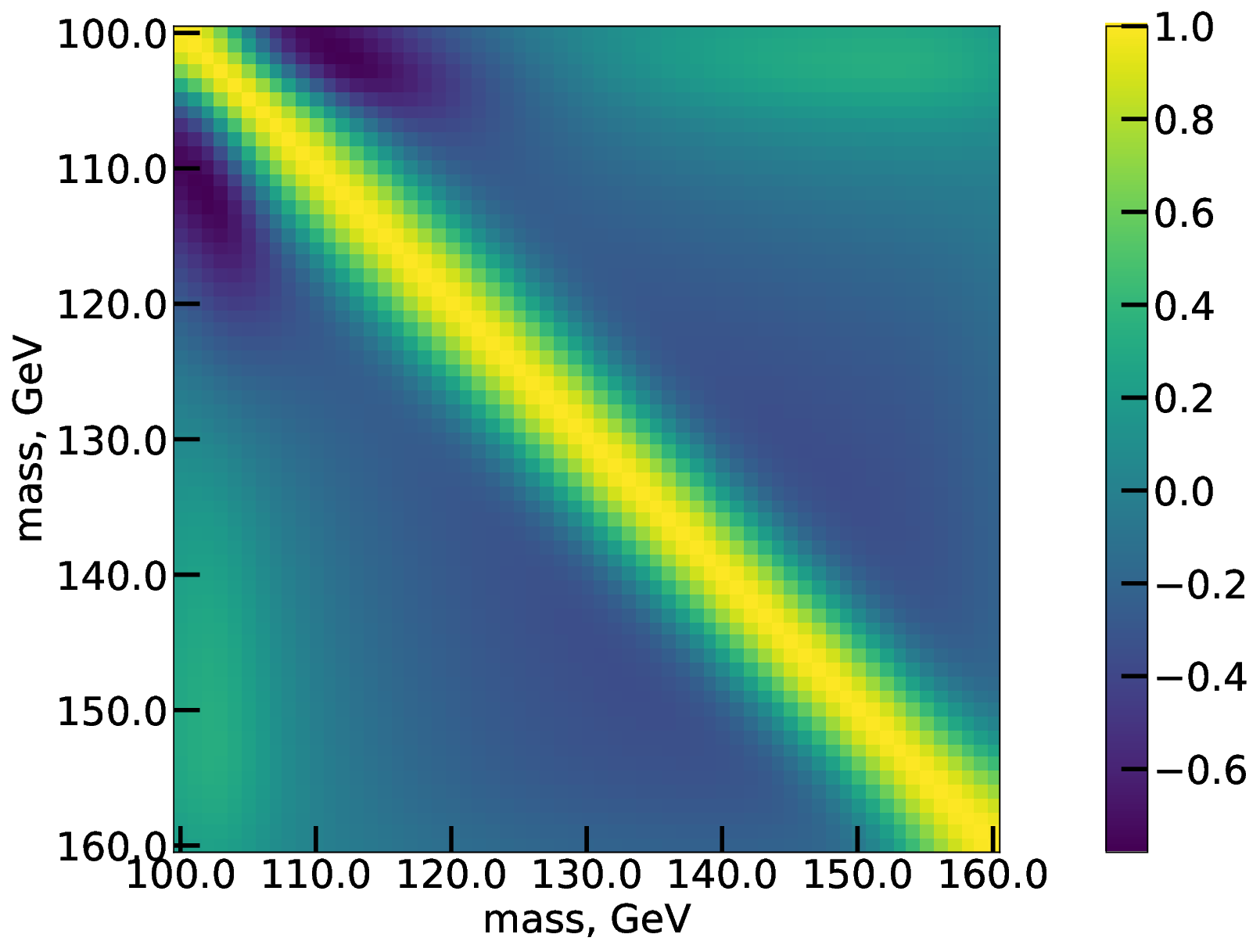}
        \caption{}\label{fig:hyy2d-covcmp_mass-mass_narrow}
    \end{subfigure} \\
    \begin{subfigure}[t]{0.45\linewidth}
        \includegraphics[width=0.94\linewidth]{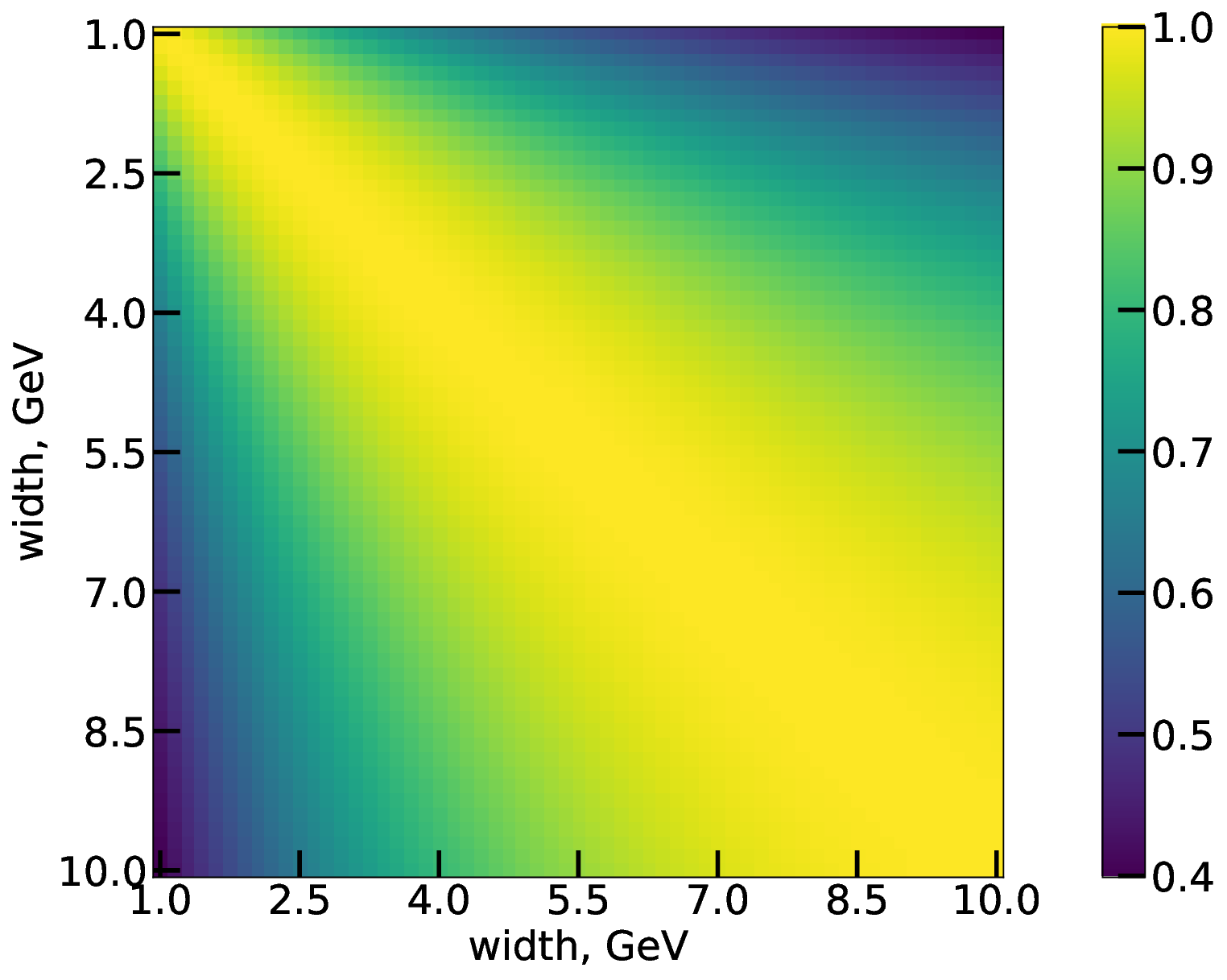}
        \caption{}\label{fig:hyy2d-covcmp_width-width}
    \end{subfigure}%
    \begin{subfigure}[t]{0.45\linewidth}
        \includegraphics[width=\linewidth]{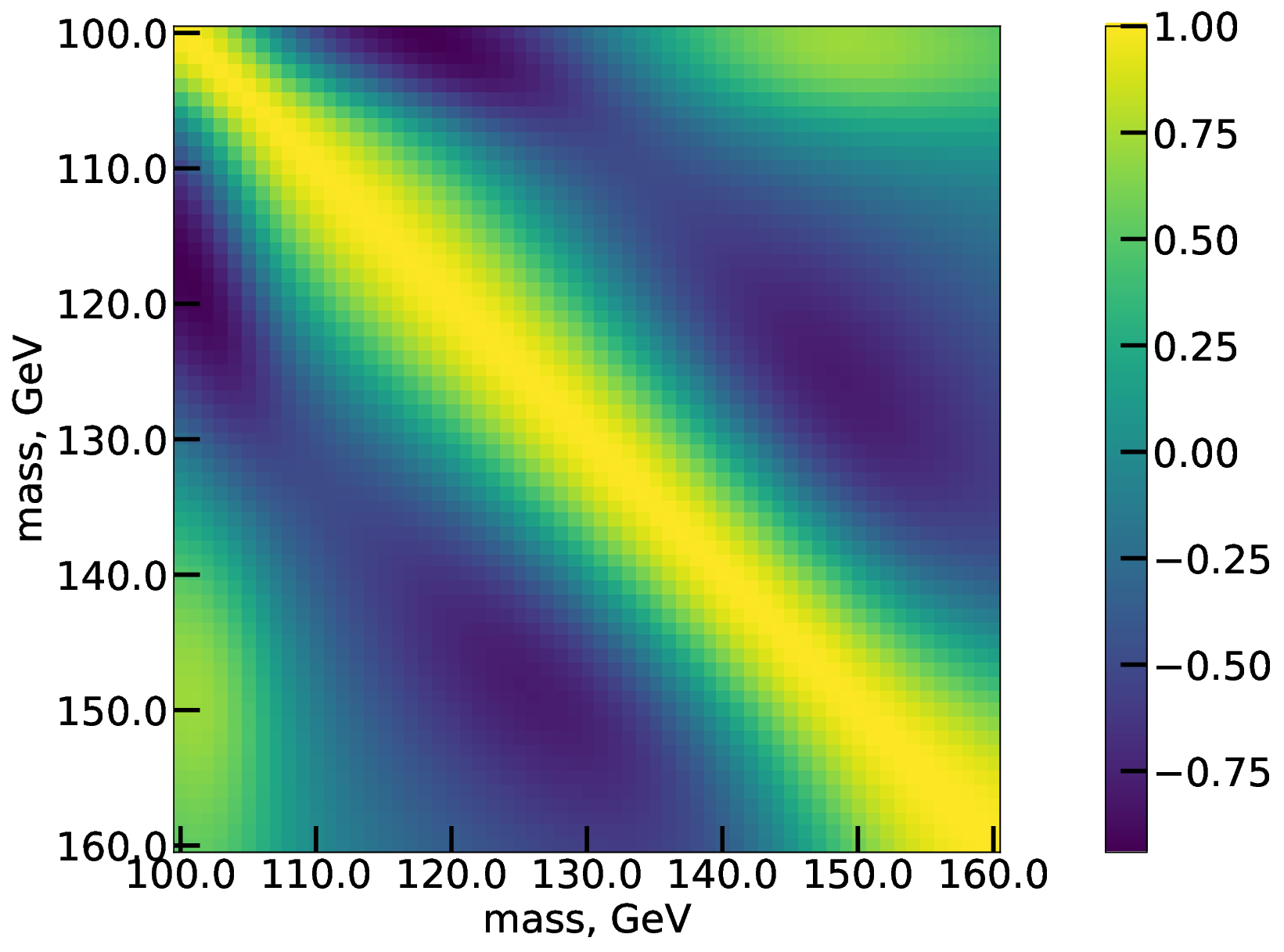}
        \caption{}\label{fig:hyy2d-covcmp_mass-mass_wide}
    \end{subfigure} \\
    \begin{subfigure}[t]{0.45\linewidth}
        \includegraphics[width=\linewidth]{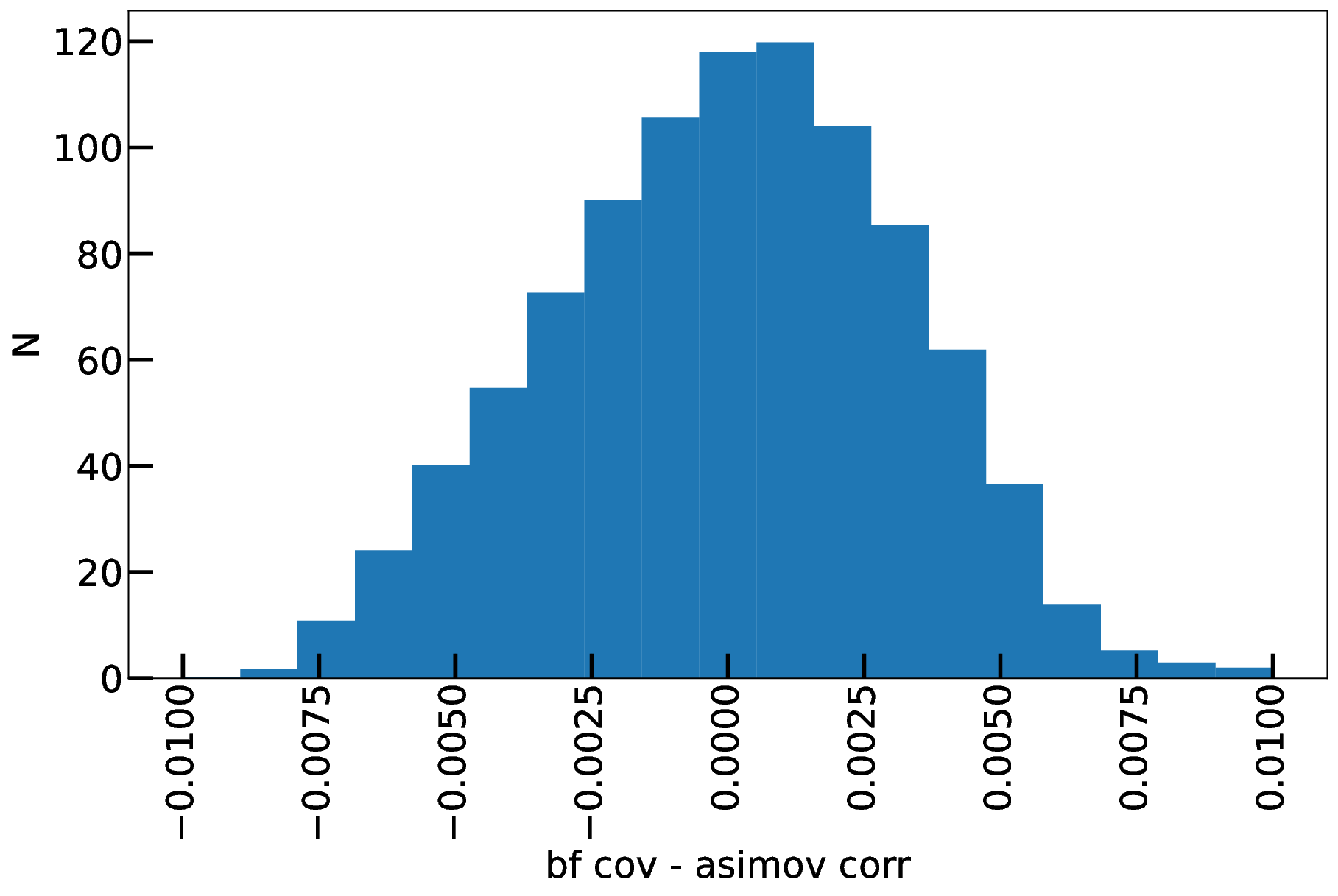}
        \caption{}\label{fig:hyy2d-covcmp_pulls}
    \end{subfigure}
    \caption{Covariance for the 2D $H\rightarrow\gamma\gamma$ model (in the 2D case a covariance between two significance surfaces).
    The covariance between the point on the first surface and 1 chosen point of mass and width ($m_s = 140~\GeV,\; \sigma_s = 2.5~\GeV$) from the second surface is shown in (\subref{fig:hyy2d-covcmp_mass-width}). The covariance between the slices of surfaces, where $\sigma_s$ was fixed for both surfaces, is shown in (\subref{fig:hyy2d-covcmp_mass-mass_narrow}) for $\sigma_s = 3.25~\GeV$ and in (\subref{fig:hyy2d-covcmp_mass-mass_wide}) for $\sigma_s = 7~\GeV$. Similarly, (\subref{fig:hyy2d-covcmp_width-width}) shows the pattern for the slices where $m_s$ is set to $130~\GeV$ for both surfaces. The histogram in (\subref{fig:hyy2d-covcmp_pulls}) shows the distribution of the values in the cells of the matrix of differences between the brute force and Asimov covariances, where $N$ denotes the absolute bin counts.}\label{fig:hyy2d-covcmp}
\end{figure}

For the covariance calculation the situation does not change if we think in terms of $M$, i.e., the expressions in eq.~(\ref{eq:sigcor_definition}) still hold. For visualization, on the contrary, it is convenient to distinguish between \textit{mass} and \textit{width} dimensions. The transition between $M_{mn}$, that preserves \textit{mass} and \textit{width} correspondingly, and $M_i$ can be constructed as follows:
\begin{align}\label{eq:2dcov_unwrap}
    \{ M_{1}, &M_{2}, \ldots \} \\
              &\Downarrow  \nonumber \\
    \{ M_{11}, \ldots , M_{1k}, &M_{21}, \ldots , M_{2k}, M_{31}, \ldots \}. \nonumber
\end{align}
In figure~\ref{fig:hyy2d-cov_full} we visualize the Asimov covariance matrix in the 2-dimensional case, which we unwrapped according to eq.~(\ref{eq:2dcov_unwrap}). We also show different projections of this covariance and compare the Asimov covariance to the brute force covariance computed with $3\cdot10^6$ toys in figure~\ref{fig:hyy2d-covcmp}. We again observe that the differences between covariances (figure~\ref{fig:hyy2d-covcmp_pulls}) are 100 times smaller than the values in the covariance matrix in figure~\ref{fig:hyy2d-cov_full}.

\begin{figure}[ht]
    \centering
    \includegraphics[width=0.48\linewidth]{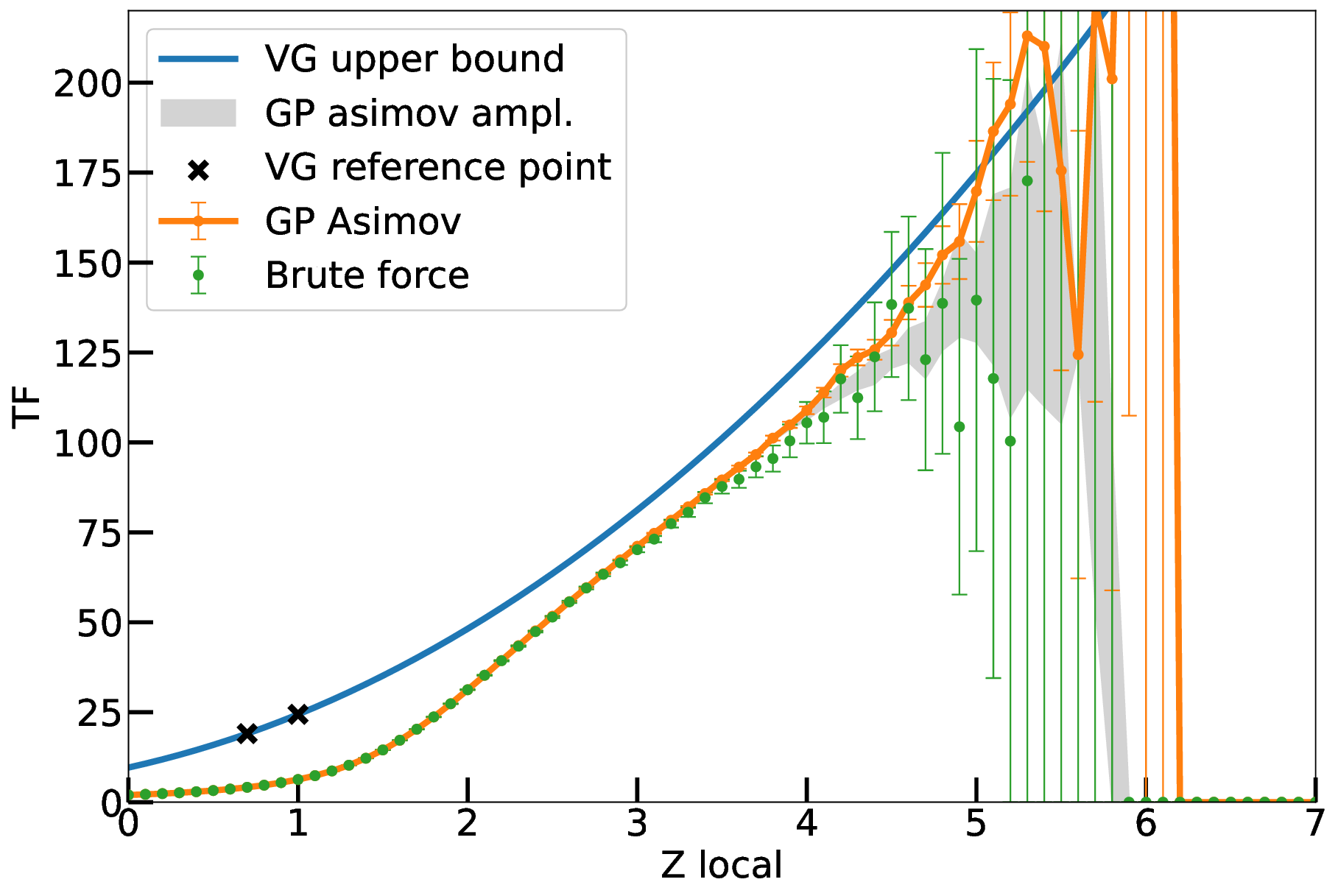}
    \caption{Trials Factor TF as a function of local significance $Z_\textrm{local}$ for the $H\rightarrow\gamma\gamma$ inspired model with 2D search region. The green dots, orange line, and gray area are explained in Fig.\ref{fig:gv-trials_factor_cmp}. The solid blue line shows the Vitells and Gross upper bound~\cite{vitells2011} extrapolated from the two reference significance thresholds ($\sqrt{0.5}$ and $1$) (black crosses).}\label{fig:hyy2d-trials_factor_cmp}
\end{figure}

The trials factors we obtained for this model are shown in figure~\ref{fig:hyy2d-trials_factor_cmp}. In this case the Vitells and Gross upper bound~\cite{vitells2011} is based on the Euler characteristic of the 2-dimensional significance surface. The upper bound calculation requires the Euler characteristic to be known at two significance levels (black crosses) before extrapolation. From the qualitative comparison of figure~\ref{fig:hyy2d-trials_factor_cmp} to figure~\ref{fig:gv-trials_factor_cmp} and figure~\ref{fig:hyy-trials_factor_cmp} it seems that for a higher dimensional search region, the Vitells and Gross upper bound becomes even more conservative.

\section{Performance analysis}

Several times in our work we mentioned that GP toys are more efficient in comparison to brute force toys.
In this section we will quantify this statement.

We chose the $H\gamma\gamma$ inspired model as a benchmark.
There were two reasons that guided our choice:
\begin{itemize}
    \item{The $H\gamma\gamma$ inspired model is more complex than the background template model,}
    \item{Adding more dimensions, according to section \ref{subsec:example_hyy2d}, only increases the number of fits per MC sample,
          but does not change the relative fitting performance.}
\end{itemize}

Fitting on a regular laptop with 4 cores and one Python process, it took us $9$~m~$30$~sec to produce $10^3$ significance curves, which is $62~000$ fits for the $61$-bin grid, or about 2 significance curves per second.

To estimate the covariance matrix with our proposed set of Asimov samples requires $(61+1)\times61=3782$ fits to obtain 61 significance curves, which took 23 seconds, or also about 2 significance curves per second.
Having the covariance, we then managed to produce $10 \cdot 10^6$ GP toys in $27$~sec, or roughly $4\times10^5$ GP toys per brute force toy.

In summary, for little additional effort and CPU time, we can create much larger samples to  estimate global p-values or average up-crossings.

\section{Conclusion}
A precise estimation of the look-elsewhere effect in searches for new resonances in high energy physics is challenging. Existing methods are overly conservative unless the observed local significance is high enough. In this paper we proposed and evaluated a new approach based on Gaussian processes. The new approach doesn't eliminate the production of Monte Carlo ``toys'', however, the computational process becomes much more efficient. We tested the approach on several models, including models with more than 1 nuisance parameter, that intentionally were constructed to have different statistical features. The trials factors calculated with our new approach and with large samples of brute force Monte Carlo simulations showed excellent agreement. As a cross-check, we reproduced the result of the Gross and Vitells paper~\cite{gross2010} that introduced the high-significance approximation based on upcrossings. An actual realization of the method and implementation of the test models was published and is available for use out of the box~\cite{gitlab-sigcorr}.

\acknowledgments{}

We would like to thank Ofer Vitells for several very helpful discussions,
in particular about the details of the statistical model in the Gross and Vitells paper.
We would also like to acknowledge the support of the ATLAS Collaboration.
This research was supported by
the European Union Framework Programme for Research and Innovation Horizon 2020 (2014--2021)
under the Marie Sklodowska-Curie Grant Agreement No.765710.

\bibliographystyle{JHEP}
\bibliography{bibliography}   

\providecommand{\href}[2]{#2}\begingroup\raggedright\begin{thebibliography}{10}

\bibitem{gross2010}
E.~Gross and O.~Vitells, \emph{Trial factors for the look elsewhere effect in
  high energy physics},
  \href{https://doi.org/10.1140/epjc/s10052-010-1470-8}{\emph{The European
  Physical Journal C} {\bfseries 70} (2010) 525–530}.

\bibitem{vitells2011}
O.~Vitells and E.~Gross, \emph{Estimating the significance of a signal in a
  multi-dimensional search},
  \href{https://doi.org/10.1016/j.astropartphys.2011.08.005}{\emph{Astroparticle
  Physics} {\bfseries 35} (2011) 230–234}.

\bibitem{davies77}
R.B.~Davies, \emph{{Hypothesis testing when a nuisance parameter is present
  only under the alternative}},
  \href{https://doi.org/10.1093/biomet/64.2.247}{\emph{Biometrika} {\bfseries
  64} (1977) 247}.

\bibitem{PhysRevLett.108.111803}
{\scshape ATLAS Collaboration} collaboration, \emph{Search for the standard
  model higgs boson in the diphoton decay channel with $4.9\text{ }\text{
  }{\mathrm{fb}}^{\ensuremath{-}1}$ of $pp$ collision data at $\sqrt{s}=7\text{
  }\text{ }\mathrm{TeV}$ with atlas},
  \href{https://doi.org/10.1103/PhysRevLett.108.111803}{\emph{Phys. Rev. Lett.}
  {\bfseries 108} (2012) 111803}.

\bibitem{CMSHyy}
{CMS Collaboration}, \emph{Search for the standard model higgs boson decaying
  into two photons in $pp$ collisions at $\sqrt{s}=7\,\textrm{TeV}$},
  \href{https://doi.org/https://doi.org/10.1016/j.physletb.2012.03.003}{\emph{Physics
  Letters B} {\bfseries 710} (2012) 403}.

\bibitem{zenodo-data}
V.~Ananyev and A.~Read, ``Datasets of fitted toy monte carlo samples for
  ``gaussian process-based calculation of look-elsewhere trials factor'':
  \url{https://doi.org/10.5281/zenodo.7861345}.''

\bibitem{gitlab-sigcorr}
V.~Ananyev and A.~Read, ``Repository containing the source code of the gaussian
  process based approach with examples:
  \url{https://gitlab.com/sigcorr/sigcorr}.''

\bibitem{gitlab-sigcorr-docs}
V.~Ananyev and A.~Read, ``Documentation and examples for {SigCorr}:
  \url{https://sigcorr.gitlab.io/sigcorr/dev/}.''

\bibitem{wilks1938}
S.S.~Wilks, \emph{The large-sample distribution of the likelihood ratio for
  testing composite hypotheses}, {\emph{The Annals of Mathematical Statistics}
  {\bfseries 9} (1938) 60}.

\bibitem{cowan2011}
G.~Cowan, K.~Cranmer, E.~Gross and O.~Vitells, \emph{Asymptotic formulae for
  likelihood-based tests of new physics},
  \href{https://doi.org/10.1140/epjc/s10052-011-1554-0}{\emph{The European
  Physical Journal C} {\bfseries 71} (2011) }.

\bibitem{lutes2004}
L.D.~Lutes and S.~Sarkani, \emph{Random Vibrations. Analysis of Structural and
  Mechanical Systems}, Butterworth Heinemann, Boston (2004).

\bibitem{cramer1967}
H.~Cramér and M.R.~Leadbetter, \emph{Stationary and related stochastic
  processes: sample function properties and their applications},  1967.

\end{thebibliography}\endgroup

\end{document}